\documentclass[showpacs,amsmath,amssymb,twocolumn,aps]{revtex4}
\usepackage{graphicx}
\usepackage{epsfig}
\usepackage{pstricks}
\usepackage{psfrag}
\usepackage{dcolumn}
\usepackage{bm}

 \def\comment#1{}
\def\mn#1{*{\marginpar{\footnotesize #1}}}
\def\mn#1{}
 \newcommand{\bol}[1]{\mbox{\boldmath$#1$}}

\begin{document}
\title{Phase Diagram of Vortices
in High-${\bf  T_c} $ Superconductors from
Lattice Defect Model with Pinning}
\author{J\"urgen Dietel}
\affiliation{Institut f\"ur Theoretische Physik,
Freie Universit\"at Berlin, Arnimallee 14, D-14195 Berlin, Germany}
\author{Hagen Kleinert}

\affiliation{Institut f\"ur Theoretische Physik,
Freie Universit\"at Berlin, Arnimallee 14, D-14195 Berlin, Germany}
\date{Received \today}
\begin{abstract}
The theory presented is based on a simple
Hamiltonian for a vortex lattice in a weak  impurity background
which includes linear elasticity
and plasticity, the latter  in the form
of integer valued fields
accounting for
defects.
Within
a quadratic approximation in
the impurity potential, we find a first-order Bragg-glass,
vortex-glass transition line showing a
reentrant behavior for superconductors with a melting
line near $ H_{c2} $. Going beyond  the quadratic
approximation by using the
variational approach of M\'ezard and Parisi
established for random manifolds, we obtain a phase diagram containing
a third-order glass transition line.  
The glass transition line separates the
vortex glass and the vortex liquid. Furthermore, we find
a unified first-order line consisting
of the melting transition between the Bragg glass and the vortex liquid phase
as well as a disorder induced first-order line between
the Bragg glass and the vortex glass phase.
The reentrant behavior of this line within the quadratic approach
mentioned above vanished. We calculate
the entropy and magnetic induction jumps over the first-order line.
\end{abstract}

\pacs{74.25.Qt, 74.72.Bk}
\maketitle

\section{Introduction}

The phase diagram of high-$ T_c $ superconductors in the $ H-T $-plane
is  dominated by the interplay of thermal fluctuations and disorder
\cite{Blatter1,Nattermann1}.
It is  believed that at low magnetic fields near $ T_c $
the vortex solid melts into a
vortex liquid (VL) via a first-order melting transition.
Prominent examples of high-$T_c$
superconductors exhibiting
a solid-liquid  melting are the
anisotropic compound
$ {\rm YBa}_2 {\rm Cu}_3 {\rm O}_{7-\delta} $ (YBCO),
 and the strongly layered
compound $ {\rm Bi}_2 {\rm Sr}_2 {\rm Ca}
{\rm Cu}_2 {\rm O}_8 $ (BSCCO).
When including weak pinning, the solid phase becomes
a  quasi-long-range
ordered Bragg glass (BG) \cite{Blatter1}.
At higher
 magnetic fields, the quasi-long-range order is destroyed
and
 there exist also a vortex glass (VG) phase.
The transition is marked by
 the disappearance of Bragg peaks in scattering data.
There is strong experimental evidence especially for BSCCO
\cite{Avraham1,Beek1}
but also for YBCO \cite{Radzyner1} that the BG-VG
transition is first
order, although in YBCO it has not been confirmed
that this is really a proper phase transition, not just a crossover.
\mn{want to say this?}
So far, the transition line has been
identified only by some magnetic anomalies in the response
to the external magnetic field.
\begin{figure}[t]
\begin{center}
\includegraphics[height=6cm,width=8cm]{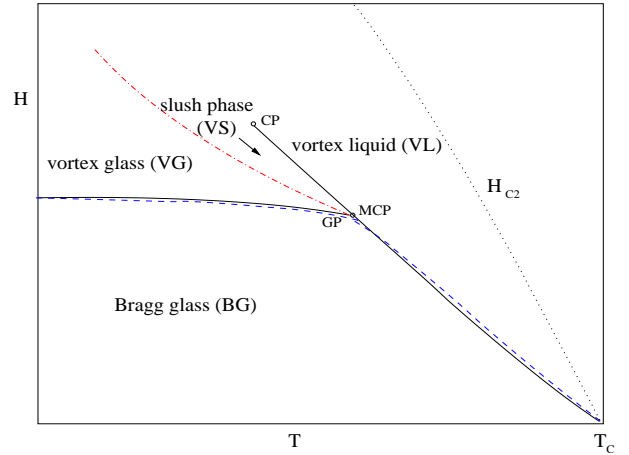}
\end{center}
 \caption{Sketch of the two possible
scenarios of the phase diagram of YBCO or similar
high-$T_c$ superconductors where the phase transition line lies near
$ H_{c2} $. The straight line corresponds to the  BG-VG, BG-VL first-order
lines with an extension of the first-order line beyond the
lower multicritical point (MCP). This is the first scenario
discussed in the text. The dashed line
corresponds to a unified BG-VG, BG-VL first-order line without a slush
phase (VS) corresponding to the second scenario.
The dashed-dotted line is the glass transition line. The point GP
is the intersection point of the glass transition line with the
BG-VG, BG-VL line.}
\end{figure}

For BSCCO it seems that the two melting lines are part of
a unified first-order transition line.
For YBCO there are two possible experimental scenarios:
First, the BG-VG and the BG-VL
 transition lines meet in a
multicritical point (MCP) \cite{Bouquet1} where the thermodynamical
character of the BG-VG line is not clear.
The first-order BG-VL melting line should
continue beyond the  multicritical point
ending in a critical point (CP) \cite{Pal1,Shibata1} where a new fluid-like
phase, the slush phase (VS), emerges (see Fig.~1).
 \mn{not understand, wo dies doch gleich
noch diskutiert wird.}

The second scenario consists of a unified
melting line \cite{Radzyner1} without a multicritical
point which  is the case for BSCCO.
We note that the experimentally realized
scenario is strongly sensitive on the doping of the superconductor
\cite{Shibata1}.

Beside the first-order transition lines,
there exits a glass transition line
between the
VG and VS phases, if the VS phase exists,
 or between the VG and VL phases, if VS is absent.
This glass transition line was predicted
by Fisher {\it et al.} in Ref.\,\onlinecite{Fisher1}, and observed
when
confirming scaling rules
for special current-voltage
characteristics across \mn{or across?} the transition line \cite{Gammel1}.
Alternatively it was proposed in Ref.~\onlinecite{Reichhardt1} that the 
glass transition  
is window glass like with no scaling. 
Some people define an irreversibility line
 beyond which  magnetization sweeps  are
no longer reversible
\cite{Nishizaki1}.
This seems to coincide
with the glass transition line.
A direct
experimental determination of the
order
of the glass transition
in the vortex system of YBCO has not
yet been possible.
For BSCCO, there is recent experimental evidence that the glass
transition line could be of second order \cite{Beidenkopf1}.
A sketch of the phase diagram for YBCO which contains
the two scenarios is shown in Fig.~1.

Beside experiments to determine the phase diagram, information comes
from computer simulations
 based on the Langevin
equation \cite{Otterlo1} or on  frustrated XY-models
\cite{Nonomura1,Olsson1,Rodriguez1}.
The Langevin simulations confirm the second phase scenario without a
slush phase, and the existence of a multicritical point on the melting line
being unclear.
In the
frustrated XY model,
the
existence of a slush phase
and of a multicritical point
are also controversial \cite{Olsson2}.
In addition,
Lidmar \cite{Lidmar1}
carried out a Monte-Carlo simulation based on a defect model
where he only obtains a first-order melting line
and a glass transition line, but not a VS phase.

 Analytic approaches are based mainly on the Ginzburg Landau model
\cite{Li1}, which is especially useful for YBCO,
the cage model \cite{Ertas1}, or the elasticity
model of the vortex lattice \cite{Giamarchi2,
Mikitik1, Mikitik2, Kierfeld1, Kierfeld2,Radzyner2, Menon1} with pinning.
The Ginzburg Landau model with pinning was analyzed recently
by Li {\it et al.} \cite{Li1} where they found a phase diagram of
 the
second  scenario, with a single first-order melting line
between the VG and VL phases as well as
between the BG and VG phases,
without an additional slush phase.
The calculation was restricted to second order in the disorder potential.
In a recent paper they
also carry out an analysis of a possible
glass transition line in the fluid phase of the Ginzburg-Landau model
where they found such a line only under a certain
disorder model  \cite{Li2}
by using  replica symmetry-breaking techniques which we
also use in this paper.
In \cite{Ertas1,Giamarchi2,Mikitik1,Menon1,Radzyner2,Kierfeld2} the
phenomenological Lindemann criterion extended to include pinning was used
in order to calculate the BG-VG and BG-VL transition lines. In
\cite{Kierfeld1,Mikitik2} defects were taken  into
account for determining the transition lines. These approaches
allow
for an explanation of both phase scenarios.

It is the purpose of this paper
to  investigate
the above phase transitions
in a defect melting model
which was recently constructed
for the study of
 defect-induced melting
of square (YBCO) and triangular (BSCCO) vortex lattices.
The model is a modification
of a simpler version
 in Ref.\,\onlinecite{GFCM2}
 which explained the
melting transition of ordinary crystals by the statistical mechanics
of defects on a hypothetical
square lattice. This model was
generalized for two-dimensional triangular crystals in
Ref.\,\onlinecite{Dietel1}. The model is Gaussian in the elastic
strains and
takes into account
the defect degrees of freedom
by integer valued gauge fields.
 The melting line
is found from a lowest-order approximation,
in which one identifies the melting point with
the intersection
of the high-temperature expansion
 of the free energy density
dominated by defect fluctuations
with the low-temperature expansion dominated by elastic fluctuations.

In this paper  we shall consider, in addition,
the effect of weak disorder on the melting
line near $ H_{c2} $. This
 will lead to a
 determination of the BG-VG and BG-VL  transition line.
The most prominent example for a high-$T_c$  superconductors
with such a  melting line is YBCO but also superconductors with a low
critical temperature $ T_c $ such as BCS type superconductors
or with a small  anisotropy factor  should have
a melting line near $ H_{c2} $. For concrete calculations, we
will restrict us in the following to the case of YBCO.

The paper will
first
review
the model and derive an effective
Hamiltonian for the vortex lattice in
the low-temperature solid phase and the high-temperature fluid  phase
without disorder.
The model has two mutually representations.
One can be evaluated efficiently in the
low-temperature phase,
the other in the high-temperature phase.
The
lowest approximation
to the former
contains only
elastic fluctuations
of the
vortex lattice without defects.
The dual representation
sums over all integer-valued
stress configurations,
which to lowest approximation
are completely frozen out.
The tranverse part of the vortex fluctuations in the high-temperature
approximation corresponds to
non-interacting three-dimensional elastic strings
where the length in z-direction is discretized with the
dislocation length as the lattice spacing \cite{Dietel2}.
It is well known, that the lower
critical dimension for an elastic string
in a random potential \cite{Halpin1} is three. This dimension
separates  the string system
in higher dimensions with two phases (a disorder dominated low-temperature
phase and a temperature dominated high-temperature phase)
from a single disorder 
dominated phase in lower dimensions. We encounter a similar
situation for the high temperature Hamiltonian in Sect.~V.
This is the reason, why we  shall have to consider higher-order
expansion terms \cite{Gorokhov1}.

We shall first expand the free energy to lowest
order in the disorder potential in Section III.
The result will be a unified melting line.
This line bends to lower magnetic fields in the direction
for decreasing  temperatures
due to the disorder, in agreement with experiments.
We obtain a remarkable reentrant behavior for this line.
We do not obtain, however,
 a good agreement with experiments at low magnetic fields. In order
to get better
agreement with
experiment and  to determine also
the glass transition line we further  calculate,
 in the solid
low-temperature phase and  in the fluid high-temperature phase, the
free energy non-perturbatively by using once
the replica-trick and further
the variational approach
set up  by M\'ezard and Parisi \cite{Mezard1} for
random manifolds and  spin-glasses \cite{Dotsenko1}.
It is based on replacing the non-quadratic part
of the replicated Hamiltonian by  quadratic one, with possible mixing of
replica fields.
A transition line from a liquid to a glass consists
within the M\'ezard-Parisi approach on a boundary in thermodynamical
space from a replica symmetric quadratic Hamiltonian
to a Hamiltonian which breaks the symmetry in the replica fields.
The best quadratic Hamiltonian in the low-temperature solid phase
is full replica symmetry broken corresponding to the BG-phase.
In the high-temperature phase
we find a region where the solution is full replica symmetric
corresponding to the VL phase. Furthermore, we find a glassy region (VG)
were the optimal quadratic Hamiltonian depends on the
form of the disorder correlation function.
By carrying a comprehensive  stability analysis
in Section VII we show that for the form of the glassy state
the kurtosis $ \kappa_1 $
defined in (\ref{880}) as functional on
the positional disorder function is relevant.
A Gaussian correlation function has kurtosis $ \kappa_1 =1  $.
For high magnetic fields near $ H_{c2} $ we obtain: \\
For $ \kappa_1 < 1$ we get a one-step replica symmetry broken solution
with a third-order phase transition line.
This corresponds to a correlation function with  flatter tip and smaller tail
than the Gaussian correlation function.
In the case $ \kappa_1 \ge 1$
we obtain a full replica symmetry broken solution. 
The free energy has the same form as in the one-step replica 
symmetry breaking case leading also 
to a third-order glass transition line.
Disorder correlation  functions with smaller tips and larger tails
than the Gaussian correlation function belong to this case.
For lower magnetic fields we obtain that
the border in the disorder function space of  one-step and 
continous replica symmetry broken solutions moves to lower kurtosis.

The VG-VL phase transition happens just at the depinning temperature of
a one-dimensional string in three dimensions subjected to impurities
\cite{Blatter1}. We calculate the free energies in
the low-temperature solid and in both high-temperature phases.
By the intersection criterion we obtain further the first-order
BG-VL and BG-VG transition line.  We do not find
an additional first-order line which separates the slush phase VS from the
vortex liquid VL. Summarizing, we obtain within our approach only
the second scenario.
In the low-temperature solid phase our analysis corresponds
to the analysis of Korshunov \cite{Korshunov1}
and Giamarchi and Doussal \cite{Giamarchi1}
using the M\'ezard-Parisi approach for the vortex lattice system
in random potentials. Because this system does not contain defects
only the Bragg glass phase can be described
correctly. Within our approach we include beside the disorder ones
also the defect degrees of freedom by integer valued fields
which are important to obtain the melting transition. This allows
us to compute the whole phase diagram for YBCO.
In this sense our theory is a direct generalization of the earlier
vortex lattice approaches in random potentials.

Beside the results above, we give in Appendix~B a derivation of some
stability theorems of saddle point solutions similar to the
theorems of Carlucci {\it et al.} \cite{Carlucci1} derived within
the large
$ N'$-limit approach of M\'ezard and Parisi \cite{Mezard1}
where $ N' $ are the number of
components of the random manifold. Because in general the number of
components of a given  random manifold is rather small it is useful
to generalize these theorems also
to the variational approach of Mezard-Parisi not existent in
literature yet.

The paper is organized as follows: \\
In Section~II we state the model of the vortex lattice with defects and
impurity degrees of freedoms.
We derive in Section~III the effective low and high-temperature Hamiltonian
of the vortex lattice without impurities.
With the help of these effective Hamiltonians we calculate in
Section~IV the BG-VG, BG-VL transition line within the second
order perturbation theory in the disorder potential.
In Section~V we introduce the M\'ezard-Parisi variational
approach. Section VI calculates the saddle point solutions of the
self-energy matrices for the variational free energy within the
M\'ezard-Parisi approach in the fluid high-temperature phase.
Section VII deals with the stability
of the calculated saddle-point solutions. Section VIII calculates the
saddle point solutions in the solid phase. In Section IX, we discuss the
phase diagram of the M\'ezard-Parisi approach for YBCO
and compare it with the experimental ones.
Furthermore jump quantities are calculated in this section.
Section X contains a summary of the paper.

\section{Model}
The partition function used here for the vortex lattice without disorder
was proposed in Ref.\,\onlinecite{Dietel2}. It is motivated by
similar melting models for two-dimensional square \cite{GFCM2}
and triangular \cite{Dietel1} crystals. Motivated by the fact that YBCO
has a square vortex lattice we restrict us here to
a discussion of the phase diagram of such type of lattice.
The generalization to triangular vortex lattices is straight forward
\cite{Dietel2} resulting only in a slight  difference in numerical values.
We
briefly summarize the important
features of the model.
The partition function of the disordered flux line lattice
can be written in the canonical form as
a functional integral
\begin{equation}
Z_{\rm fl}\!= \! \int  {\cal D}[u_i,\sigma_{im},n_i] e^{-
 \left(H_0[u_i,\sigma_{im},n_i]+H_{\rm dis}[u_i]\right)/k_BT
} \,,  \label{5}
\end{equation}
where
\begin{widetext}
\begin{align}
&\frac{ H_0[u_i,\sigma_{im},n_i]}{k_BT} =
 \sum_{{\bf x}}\frac{1}{2 \beta}
 \Bigg[\sum_{i<j}\sigma _{ij}^2 +\frac{1}{2}\sum_i \sigma _{ii}^2
-
\Big(\!\sum_i \frac{\overline{\nabla}_i}{\nabla_i} \sigma _{ii}\! \Big)
 \frac{c_{11}-2c_{66}}{4(c_{11}-c_{66})}\Big(\!\sum_i
\frac{\overline{\nabla}_i}{\nabla_i}  \sigma _{ii}\! \Big)
+\sum_i
\sigma_{i3}
\frac{c_{66}}{c_{44}}\sigma_{i3}
 \Bigg] \nonumber \\
& -
2 \pi i \sum_{{\bf x}} \left(
 \sum_{i,m} \sigma_{im}
  \nabla_{m}  u_i+  \sum_{i \le j}
\sigma_{ij}
  N_{ij}\right)
 \label{2}
\end{align}
\end{widetext}
is the canonical representation of elastic and plastic energies
summed over the lattice sites
$ {\bf x} $ of a three-dimensional
lattice,   and
 $ \sigma_{ij} $ where $ \sigma_{21} \equiv \sigma_{12} $ 
are stress fields  which are canonically
conjugate to the distortion  fields
 \cite{GFCM2}.
The subscripts $ i,j $ have
the values $ 1,2$,  and $l, m,n$ run from  $  1$ to $3$.
The parameter $ \beta $
 is
proportional to the inverse temperature,
 $ \beta \equiv
a^2 a_3 c_{66}/k_B T (2\pi)^2  $,
where $a$ is the transverse distance of neighboring vortex lines,
and
 $ a_3 $ is the persistence length
along the dislocation lines introduced in
 Ref.\,\onlinecite{Dietel2}.
Note that $a_3$ is assumed to be
independent
on the disorder potential on the average.
The volume
of the fundamental cell $ v $ is equal to
  $ a^2 a_3 $ for the square lattice.

The
matrix $ N_{ij}({\bf x}) $ in Eq.~(\ref{10})
is a discrete-valued local defect matrix composed of integer-valued
defect gauge fields $n_1,n_2$. It depends on the lattice symmetry
\cite{Dietel1}. For a square vortex lattice it
is given by
\begin{eqnarray}
N_{ij} \!\!&=& \!\!\left(
 \begin{array} {cr}
n_1 & n_2 \\
n_2   & -n_1
\end{array}\right).
\label{10}
\end{eqnarray}

The lattice derivatives $ \nabla_m $ and their conjugate counterparts
$ \overline{\nabla}_m $
are the lattice differences for a cubic three-dimensional crystal.
In the xy-plane they are defined by
\begin{eqnarray}
 \nabla_i f({\bf x}) &\equiv& [f({\bf x})- f({\bf x}-a {\bf e}_i)]/a ,\nonumber \\
~~ \overline{\nabla}_i f({\bf x}) &\equiv&
 [f({\bf x}+a {\bf e}_i)- f({\bf x})]/a
\label{@}\end{eqnarray}
for  a lattice function $ f({\bf x}) $,
where
 $ {\bf e}_i $
are
 unit vectors to the
nearest neighbors in the plane. The corresponding derivatives
in $z$-direction are defined similarly.
We have suppressed the spatial arguments
of the elasticity parameters, which are
functional matrices
$ c_{ij}({\bf x},{\bf x}')\equiv
 c_{ij}({\bf x}-{\bf x}') $.
Their precise forms were first calculated by Brandt \cite{Brandt1}
and  generalized
in Ref.~\cite{Dietel2} by taking into account thermal softening relevant
for BSCCO.

The second term in the exponent
of (\ref{5})
\begin{equation}
 H_{\rm dis}[u_i]
= \sum_{{\bf x}} V({\bf x}+ {\bf u}),
 \label{4}
\end{equation}
accounts for disorder.
The measure of the functional integral is
\begin{eqnarray}\!\!\!\!
\!\!\!\!\!\!\!\!
\!\!\!\!\!\!\!\!
 \int {\cal D}[u_i,\sigma_{im},n_i]\!=\!
\rm {det} \left[\frac{c_{66}}{4(c_{11}\!-\!c_{66})}\right]^{1/2}
\!\!\!\!\!\rm{det} \left[\frac{1}{2\pi\beta}\right]^{5/2}~~~~~~\nonumber \\
~~\times \!
\left\{\! \prod_{{\bf x}}\!
 \Bigg[ \prod_{i\leq m} \int_{-\infty}^\infty d\sigma_{im}\!\Bigg]
 \Bigg[\!\prod_{j}\sum_{n_{j}({\bf x})=-\infty}^{\infty} \Bigg]
 \Bigg[\!\int_{-\infty}^\infty\frac{d {\bf u}}{a}
 \Bigg]\! \right\}. ~ \label{1}
\end{eqnarray}
The disorder potential $ V({\bf x}) $ due to pinning is assumed
to possess the
Gaussian short-scale correlation function
\begin{eqnarray}
 \overline{V({\bf x}) V({\bf x}')} & = &  \Delta(x_i-x_i')
\delta_{x_3,x_3'}  \label{11} \\
 & = &
d(T)\,  a_3 \,  \frac{\phi_0^4 \,\xi^3_{ab}}{\tilde{\lambda}_{ab}^4}
K(x_i-x_i') \,  \delta_{x_3,x_3'}      \nonumber
\end{eqnarray}
where $ K(x_i-x_i') \approx 1/(\xi')^2  $ for
$ |{\bf x}-{\bf x}'|<\xi' $, and is zero elsewhere,
and $ \phi_0 $ is the magnetic flux quantum $ \phi_0=hc/2e $. The parameter
$ \xi' $ is the correlation length of the impurity potential which
is similar to the coherence length $ \xi_{ab} $
in the $xy$-plane.
$ \tilde{\lambda}_{ab}=\lambda_{ab}/(1-b) $ is the screened
penetration depth in the $xy$-plane \cite{Mikitik1}.

The temperature dependence of the parameter $ d(T) $ is mainly
due to the temperature dependence of the correlation length and the
pinning mechanism where we discuss in the following the
$ \delta
T_c $-pinning or $ \delta_l $-pinning mechanisms \cite{Blatter1}.

Both pinning mechanisms are extensively discussed in the review of
Blatter {\it et  al.} \cite{Blatter1}.
We just mention that the $ \delta T_c $-pinning mechanism has its origin
in fluctuations in $ T_c $ in the Ginzburg-Landau  free energy and the
$ \delta l $-pinning mechanism is due to fluctuations in the
mean free path coming from fluctuations in the
impurity density.
The parameter $ d(T) $ is different for both pinning mechanisms
\cite{Blatter1}:
\begin {eqnarray}
 d(T)   & = &     d_0  (1-T/T_C)^{-1/2}  \quad  {\rm for}
\quad   \delta T_c-{\rm pinning}  \,,    \label{12}  \\
  d(T)  &  = &   d_0  (1-T/T_C)^{3/2}  \quad  \; \; {\rm for} \quad
\delta l-{\rm pinning}   .     \label{13}
\end{eqnarray}
The correlation functions for both mechanisms
 can be derived
in Fourier space
by
taking into account
the order parameter shape of a single vortex \cite{Blatter1}.
This is of long range, resulting in a divergence of the
Fourier transformed disorder correlation function $ \hat{K}(q) $
at $ q=0 $
for the $ \delta T_c $-pinning mechanism. This divergence
is regulated
 for a vortex
in a lattice by omitting the
regime
$ q \lesssim  1/a $
because the
order parameter of the superposition of non-cut off
single vortex order parameters
on the lattice would otherwise scale with the system size.
We shall see below in Section~VII that
this is the momentum region of the disorder correlation
function  which determines mainly the form of the
free energy in the fluid phase near the glass transition line and thus
the order of the glass transition.
Other correlation mechanism as for example
screening of impurities are not taken into account in these
single vortex disorder correlation  functions.
The screening of impurity potentials is important because the
nearest neighbor distance between impurities is typically
of the same size as the coherence length $ \xi_{ab} $ \cite{Blatter1}.

All this leads us to use
in the
 calculations to follow an effective
disorder correlation function with the  Fourier transform
\begin{equation}
 {\hat K}(q ) = 2 \pi \exp(- {\xi'}^2 q_i^2/2 )
 \qquad 
  \label{15}
\end{equation}
leading also to an exponentially  vanishing of the disorder
correlation function  in real space.
The advantage for using this effective correlation function
is that one gets simple analytical formulas in the
calculations. The parameter
$ \xi' $ in (\ref{15}) is an effective correlation length
which can also include for example screening effects of the
impurities in the  $ \delta_l $-pinning case.
The approximation (\ref{15})
leads to well  known approximations for the temperature
softening of quantities which use the disorder correlation functions
as an input as for example the temperature softening of
the coherently time-averaged pinning energies
\cite{Blatter1}.

In the following sections, we come back
to the more general case without the assumption (\ref{15}) for the disorder
correlation function
especially in Section~VII where we show that
the order of the glass transition line depends strongly on the
form of the correlation function. The free high-temperature energy formulas
(\ref{32}), (\ref{230}), (\ref{280}), (\ref{590}) and (\ref{1080}), are
valid irrespective of the form of the disorder correlation function.
In the low-temperature regime, the form of the energy expression
cannot be find out for general correlation functions.
In this case we restrict
us to the effective disorder potential  (\ref{15}) where
in contrast to the glass transition line,  the final energy expression
should not change much when changing the disorder potential.

\section{Partition function of solid and fluid phases
 for \boldmath $  V=0 $ }
In this Section we determine the partition function of the low
temperature phase  (solid phase) and the high-temperature phase
(fluid phase) for $ V=0 $. This was done before in Ref.\,
\onlinecite{Dietel2}. Here, we give similar expressions
which are appropriate for
calculating correlation functions of vortex displacements useful
for a discussion of the disorder problem.

For the low-temperature limit of the partition function in
 (\ref{10}) we first integrate out the stress fields $ \sigma_{ij} $.
Then the low-temperature part of the
partition function corresponds to the defect configuration $ n_i =0 $.
This results in a fluctuating part of the form
\begin{equation}
Z_{\rm fl} = {\cal N} \prod_{{\bf x},i}\left[
 \int_{-\infty}^\infty\frac{u_i({\bf x})}{a} \right]\exp\left[-\frac{1}{k_B T}
H_{0}[u_i]\right]           \label{18}
\end{equation}
with the low-temperature Hamiltonian
\begin{align}
&  H_0[u_i]=
H_{\rm T \to 0}[u_i]  =  \frac{v}{2} \sum_{{\bf x}}
 (\overline{\nabla}_i u_i) (c_{11}-2c_{66}) (\overline{\nabla}_i u_i)
  \nonumber  \\
&   +\frac{1}{2}( \nabla_i u_j + \nabla_j u_i)\,  c_{66} \,
(\nabla_i u_j + \nabla_j u_i) +
  (\nabla_3 u_i) \, c_{44} \, (\nabla_3 u_i)     \nonumber \\
& = \frac{v}{2} \sum_{{\bf x}} (\nabla_i  u_L)\, c_{11} \,
(\nabla_i u_L) +
(\nabla_3 u_L) \, c_{44} \, (\nabla_3 u_L)  \nonumber \\
& + (\nabla_i  u_T) \, c_{66} \, (\nabla_i u_T) +
(\nabla_3 u_T) \, c_{44} \, (\nabla_3 u_T)
\label{19}
\end{align}
and the normalization factor $ {\cal N}=1$.
Here $ {\bf u}_L = {\bf P}_L {\bf u} $ is the longitudinal part of the
displacements
where the projector $ {\bf P_L} $ is given by
$ (P_L)_{jk} \equiv  - (1/\sqrt{|\nabla^2_i|}) \nabla_j \otimes
(1/\sqrt{|\nabla^2_i|}) \overline{\nabla}_k $.
The transversal part of the displacements is then given by
$ {\bf u}_T = {\bf P}_T {\bf u} \equiv {\bf u}-{\bf u}_L $.
 The corrections to the fluctuating part of the free energy $ -
\ln(Z_{\rm fl})/k_B T $
in the low-temperature expansion is exponentially vanishing
with an exponent proportional to $ -1/k_B T $ \cite{GFCM2}.

For the high-temperature limit of the partition function (\ref{5})
we carry out first the sum over the defect fields $ n_1 $, $n_2 $.
By a redefinition of the stress fields
$ \sigma_g = (\sigma_{11}+ \sigma_{22}) $ and
$ \sigma_u= (\sigma_{11}- \sigma_{22}) $
we obtain that $ \sigma_{12} $ and $ \sigma_{u} $ can only have integer
numbers. The lowest-order terms in the high-temperature expansion
of the partition function (\ref{5}) for $ V=0 $
corresponds to $ \sigma_{12}= \sigma_{u} =0 $. After carrying out
the integrals over the stress fields $ \sigma_g $
and $ \sigma_{i3} $
we obtain a partition function of the form (\ref{18}) with

\begin{align}
 &  H_0[u_i]=
H_{\rm T \to \infty }[u_i]  =    \frac{v }{2} \sum_{{\bf x}}
 (\overline{\nabla}_i u_i) (c_{11}- c_{66}) (\overline{\nabla}_i u_i)
  \nonumber \\
&  \qquad \qquad +(\nabla_3 u_i) \, c_{44} \, (\nabla_3 u_i)  \nonumber \\
& =  \frac{v}{2} \sum_{{\bf x}}
 (\nabla_i  u_L)\, ( c_{11}- c_{66})
(\nabla_i u_L) +
(\nabla_3 u_L) \, c_{44} \, (\nabla_3 u_L)  \nonumber \\
& \qquad \qquad +
(\nabla_3 u_T) \, c_{44} \, (\nabla_3 u_T)                 \label{25}
\end{align}
and $ {\cal N}=  1/(4 \pi \beta)^N $.
Similar as in the case of the low-temperature expansion one can show
that corrections to the fluctuating part of the free energy $ -
\ln(Z_{\rm fl})/k_B T $ due to non-zero terms in the stress fields
 $ \sigma_{12} \not=0 $ or $ \sigma_{u} \not=0 $
in the high-temperature expansion are exponentially vanishing
with an exponent proportional to $ -k_B T $ \cite{GFCM2}.

Expression (\ref{25}) shows the remarkable fact that the transverse part
of the high-temperature Hamiltonian (\ref{25})  is
effectively one-dimensional with a non-zero dispersion only in z-direction.
This results in diverging thermal fluctuations in $ {\bf u}_T $. In contrast
to this we obtain for that  part of the Hamiltonian (\ref{25})
corresponding to longitudinal fluctuations
an effectively three-dimensional Hamiltonian
as in the low-temperature case (\ref{19})
with finite temperature fluctuations. This can be
better understood by the fact that only the transverse fluctuating part
of the vortices couples to the defect fields while the
longitudinal part is still not effected by them
\cite{Dietel2,Labusch1,Marchetti1}. The reason is that the flux lines in
a vortex lattice cannot be broken which means that defect lines
are confined in the plane spanned by their Burger's vector
and the magnetic field. In conventional crystals we do not have such a
constrained \cite{GFCM2}. It then clear that the large thermal
fluctuations of the transverse part results
in a destruction of the long range order in the sense that Bragg peaks
are vanishing in the fluid phase.

Summarizing, with the help of the stress representation (\ref{2}) we
obtained the lowest-order Hamiltonians for the solid (\ref{19})
and the fluid phase (\ref{25}).
We saw further that the higher-order corrections to this lowest-order
results corresponds to integer-valued
defect contributions $ n_i \not=0 $ in the solid phase,
signals for the liquid,
and integer valued stress contributions $ \sigma_{12}\not=0 $ or
$  \sigma_{u} \not=0$ in the fluid phase, signals typical for a solid.
In the following we restrict
us to  the lowest-order Hamiltonians (\ref{19}) for the solid phase
and (\ref{25}) for the fluid phase to discuss disorder corrections
in both phases.

\section{Quadratic approximation in
Disorder Strength}

To lowest non-vanishing order in the disorder potential
$ V $ we obtain for the
first non-vanishing term in the free energy
$ F=-k_B T \ln(Z) $ a term proportional $ V^2 $ given by
\begin{eqnarray} && \! \!\!\!\!\!\!\!
F_{{\rm fl},V^2} =    - \frac{1}{2 k_B T} \!\!\!\!\!\!\!\!\!\!\!\!
\label{30} \\
& &\!\!\!\!\!\!\!\!\!\times \bigg( \sum_{{\bf x},{\bf x}'}
\overline{\langle V({\bf x} +{\bf u}) V({\bf x'} +{\bf u}) \rangle}\! -  \!
\overline{\langle V({\bf x} +{\bf u}) \rangle
\langle V({\bf x'} +{\bf u}) \rangle}  \bigg) \,.
 \nonumber \end{eqnarray}
Note that the dimension of $ \Delta $ is $ (k_B T)^2 $ (\ref{11}).

 We restrict us to the diagonal summands $ {\bf x}= {\bf x}' $ where
non-diagonal terms results in corrections only to the low-temperature
expansion of $ F_{{\rm fl},V^2} $ being a factor
$ (\langle u^2 \rangle+ \xi'^2)^{1/2}/ a $ smaller than the diagonal terms.
The calculation
can  be most easily done by working in the Fourier representation

By using (\ref{19}) and (\ref{25})  we obtain for the low- and high-temperature
part of the free energy
\begin{eqnarray}
 F^{T \to 0}_{{\rm fl},V^2}  & = &    \frac{-N}{2 (k_B T)}
\int  \frac{d^3 q}{(2\pi)^3}
\overline{V({\bf q})V(-{\bf q})}    \nonumber  \\
& &
\times \left(1-  \exp\left[- {\bf q} \cdot \langle {\bf u}  {\bf u}
\rangle_{T \rightarrow 0} \cdot {\bf q} \right]\right)
 \,,     \label{31}    \\
  F^{T \to \infty}_{{\rm fl},V^2} & = &   \frac{-N}{2 (k_B T)}
 \int  \frac{d^3 q}{(2\pi)^3}
\overline{V({\bf q})V(-{\bf q})}      .             \label{32}
\end{eqnarray}
For the determination of the
melting line only the second term in the bracket
of $ F^{T \to 0}_{{\rm fl},V^2}$ is relevant
because of a cancellation when determining the intersection of the high and
low-temperature expansions of the free energy.

After carrying out the
momentum integral and disorder averaging we obtain for this term
$ N {\cal D}  k_B T/2  $ with the disorder
constant $ {\cal D} $ defined by
the help of the generalized disorder constants
 \begin{align}
& D_{0}(2 \langle u^2 \rangle)   =
 d(T)
 \frac{a_3 }{(k_B T)^2 } \frac{ \phi_0^4 \,
\xi_{ab}^3 }{\tilde{\lambda}^4_{ab}}
\!  \int\!   \frac{d^2q}{ (2 \pi)^2}\,  \hat{K}(q) \,
   e^{-\frac{q^2}{2} \langle u^2 \rangle}  ,
                       \label{33}    \\
& D_{ \infty}(q)  =
 d(T)
 \frac{a_3}{(k_B T)^2 } \frac{\phi_0^4 \,
\xi_{ab}^3 }{\tilde{\lambda}^4_{ab}}
    \frac{\hat{K}^2(0)}{2\pi}  \bigg/ \frac{d}{
d (q^2/2)} \hat{K}(0)  \label{34}
\end{align}
where  $ {\cal D}(2 \langle u^2 \rangle)
= D_{0}(2 \langle u^2 \rangle) $. Note that we have  $ {\cal D}(0) =
{\cal D}_{\infty} (0)={\cal D}_{0} (0) $ for
the Gaussian correlation function (\ref{15}). For this correlation function, we obtain
\begin{equation}
  {\cal D} (2 \langle u^2 \rangle)
  \approx  d(T)
 \frac{a_3}{(k_B T)^2 } \frac{\phi_0^4 \, \xi_{ab}}{\tilde{\lambda}^4_{ab}}
  \frac{ \xi_{ab}^2 }{[(\xi')^2+  \langle u^2 \rangle ]}       \,.
 \label{36}    \end{equation}
Furthermore, we define the corresponding disorder correlation lengths
by
\begin{eqnarray}
\frac{1}{{\xi'}^2_{0}} & = &
\frac{1}{(2\pi)} \int d^2 q \, \hat{K}(q)\bigg/\hat{K}(0)
   \,,  \label{37}  \\
\frac{1}{{\xi'}^2_{\infty}} & = &
\hat{K}(0)\bigg{/} \frac{d}{d(q^2/2)}\,  \hat{K}(0)      \label{38}
\end{eqnarray}
with $ \xi'^2  = \xi'^2_{\infty}=  \xi'^2_{0} $ for the
Gaussian  correlation function (\ref{15}).
In the following, we carry out the calculation of the
free energies explicitly for the
Gaussian correlation function where we use $ {\cal D} $ and $ \xi' $ without indices.
As mentioned in the introduction our
final results in this section and also
for the fluid phase in the M\'ezard-Parisi approach
are more general valid without restrictions on the
disorder correlation functions.

By recalling  the results for $ Z_{\rm fl} $
without disorder \cite{Dietel2}
with the low-temperature Hamiltonian (\ref{19})
we obtain
\begin{equation}
Z^{T \to 0}_{\rm fl,0}= \left(\frac{a_3}{a} \right)^{2N}
\frac{1}{{\rm det} \left[(2 \pi \beta){c_{44}}/{c_{66}}
\right]} \;
e^{- N \sum_{i \in\{1,6\}} l_{ii}}      ,           \label{40}
\end{equation}
and for the high-temperature part (\ref{25})
\begin{equation}
Z^{T \to \infty}_{\rm fl,0}=
\left(\frac{a_3}{a} \right)^{2N} \frac{1}{2^N}
\frac{1}{{\rm det}\left[(2\pi \beta)^2 {c_{44}}/{c_{66}}
\right]}  \,  e^{- N h}      \label{45}
\end{equation}
with
\begin{align}
& l_{ii}=\frac{1}{2} \frac{1}{V_{\rm BZ}} \int_{\rm BZ} d^2k dk_3 \,
\ln \left[\frac{c_{ii} a_3^2}{c_{44}}
 K^*_j K_j  + a_3^2 K^*_3 K_3
\right]  , \nonumber \\
& h =\frac{1}{2} \frac{1}{V_{\rm BZ}} \int_{\rm BZ} d^2k dk_3 \,
\ln \left[1+\frac{c_{11}-c_{66}}{c_{44}} \,
\frac{K^*_j K_j}{ K^*_3 K_3}
\right]   \label{50}
\end{align}
where $ K_m $ is the eigenvalue of $ i \nabla_m $.
The $ k,k_3  $-integrations in (\ref{40}) run over the Brioullin zone
of the vortex lattice
of volume $ V_{\rm BZ}= (2 \pi)^3 /v$.
According to the
intersection criterion we equate (\ref{40})
and (\ref{45})
and obtain the equation for the temperature
\cite{Dietel2}
\begin{equation}
\frac{k_B T }{v}
\frac{1}{{\rm det}^{1/N}[c_{66}]}
= \frac{ e^{-(l_{11}+l_{66})+h -{\cal D}/2 }}{\pi} \,.
              \label{55}
\end{equation}
The solution determines the first-order BG-VG, BG-VL transition line
with disorder.
The disorder enters
the equation via the disorder function  $ {\cal D} $.
Analytic expressions can be obtained by taking into account that
 $ c_{66}, c_{44} \ll c_{11} $. This implies
that we can neglect $ h $ and $ l_{11} $ in (\ref{40}).
\begin{figure}[t]
\begin{center}
\includegraphics[height=8cm,width=8.5cm]{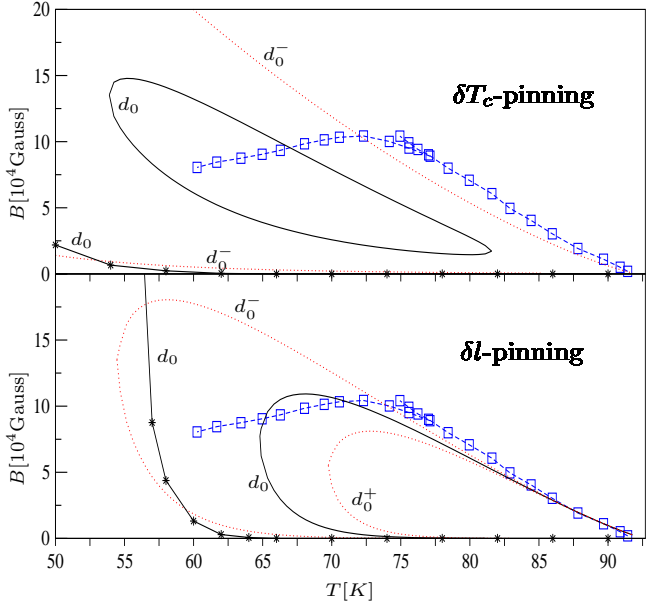}
\end{center}
 \caption {Unified BG-VG, VG-VL transition line $ B_m(T) $ (\ref{65})
as a function of the temperature. The curves in the
upper figure are  calculated for
$ \delta T_c $-pinning mechanism (\ref{12}), the lower for
$ \delta_l $-pinning mechanism (\ref{13}). The parameter $ d_0 $
of the solid curves are chosen such that we get a good
fitting to the experimentally determined VG-VL line by Bouquet {\it et al.}
\cite{Bouquet1}
(dashed curve with square points)
for both pinning mechanisms. The dotted curves are
variations from these best fitting curves given by
disorder parameters
$ d_0^{\pm}=(1 \pm 1/2) \,d_0 $
where $ d_0 $ are the disorder parameters of the solid curves
of both mechanisms given by $ 2\pi d_0 \, \xi^2_{ab}/{\xi'}^2
=8.5 \cdot 10^{-8} $ ($ \delta T_c $- pinning)
and $ 2\pi d_0 \, \xi^2_{ab}/{\xi'}^2
 =1.01 \cdot 10^{-6} $ ($\delta l $- pinning).
 The solid curves with the stars are calculated by solving (\ref{55})
 with elastic moduli in the range $ b \lesssim 0.2 $ with $ d_0 $
 given above.
}
 \vspace*{0cm}
\end{figure}

\begin{figure}[t]
\begin{center}
\includegraphics[height=6cm,width=8.5cm]{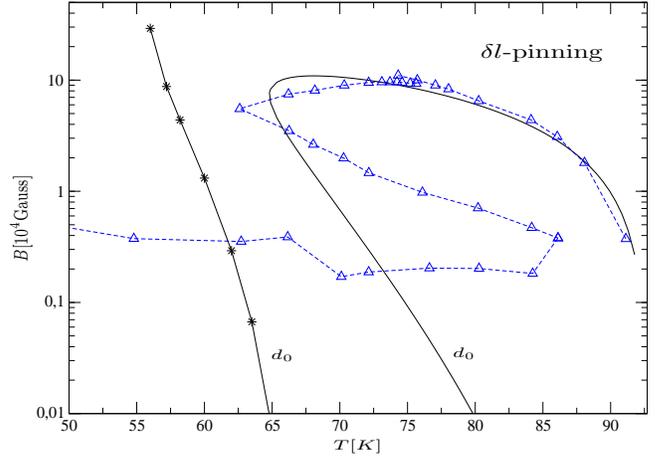}
\end{center}
 \caption{Logarithmic plot of the BG-VG, BG-VL first-order line
$B_m(T) $ (\ref{65}). The experimental points (triangles) correspond
to the experiment of Pal {\it et al.} \cite{Pal1} showing a reentrant
BG-VG line.
The theoretical determined solid curves
are  derived from (\ref{65}). The pure solid curve ($ b \gtrsim 0.5$)
and the solid curve with the
stars ($ b \lesssim 0.2 $) correspond to the lower curves
in Fig.~2 ($ \delta l $-pinning).  The theoretical curves are  calculated
with the disorder parameter $  2\pi d_0 \, \xi^2_{ab}/{\xi'}^2
=1.01 \cdot 10^{-6}  $ of Fig.~2.}
\vspace*{0cm}
\end{figure}

Brandt \cite{Brandt1} determined the elastic constants for two different
regimes $ b \lesssim 0.2 $ and $ b \gtrsim 0.5 $ where $ b= B/H_{c2}(T) $.
We shall  see below that
 for YBCO
we have to
determine  (\ref{55}) in both regimes
to find
the entire relevant part of the BG-VG and BG-VL line.
The most important part, however,
 lies in the regime
$ b \gtrsim 0.5 $ which will now be treated  explicitly.
In this regime the elastic moduli
$ c_{44} $ and $ c_{66} $
are given by  \cite{Brandt1}
\begin{eqnarray}
 c_{66} \! &  \!= \!&  \!
0.71 \, (1-b) \frac{B \phi_0 }{64 \pi^2 \tilde{\lambda}_{ab}^2}
\,, \label{58} \\
 c_{44}\! & \! = \!&  \!
\frac{B^2}{4 \pi(1\!+ \! \tilde{\lambda}_{c}^2 k^2\! + \!
\tilde{\lambda}_{ab}^2 k_3^2)}+
\frac{B \phi_0}{16 \pi^2 \tilde{\lambda}_c^2} .
\label{60}
\end{eqnarray}
$ \tilde{\lambda}=\lambda/(1-b)^{1/2} $ is the screened penetration depth
calculated from the penetration depth $ \lambda $.
In $ xy $-direction,
we denote it by $ \lambda_{ab} $, and in
$ z $-direction by $ \lambda_c $.
For YBCO we have \cite{Kamal1} $ \lambda(T)=
\lambda(0)[1-(T/T_c)]^{-1/3}$, $ \xi_{ab}(T)=
\xi_{ab}(0)[1-(T/T_c)]^{-1/2} $.

For later use, we define the Lindemann parameter \cite{Dietel2}
\begin{eqnarray}
 c_L^2\! & = & \!\frac{a_3^2 }{a^2 v}  \frac{k_B T}{V_{\rm BZ}}
\int_{\rm BZ} \! \!d^2k dk_3 \frac{1}{c_{44}}\sum_{i=1,6}  \!
\frac{1}{\frac{c_{ii} a_3^2}{c_{44}}
  K^*_j K_j  + a_3^2 K^*_3 K_3 }   \nonumber \\
& \approx &
\frac{k_B T_m }{ 4  \left[ c_{44}(\frac{K_{\rm BZ}}{\sqrt{2}},0)\,
c_{66}(\frac{K_{\rm BZ}}{\sqrt{2}} ,0)\right]^{1/2} a^3}\,.  \label{62} \
\end{eqnarray}
given by $ c_L^2=
\langle u^2 \rangle_{T \to 0} /a^2 $ where the average is
taken with respect to the
low-temperature Hamiltonian (\ref{19}) representing the elastic energy of
the vortex lattice.  $ K_{\rm BZ} $ is the boundary of the
circular Brillouin zone $ K_{\rm BZ}^2=4 \pi B/\phi_0 $.
For YBCO, we obtain \cite{Dietel2}
$ c_L\approx 0.18 $ on the melting line
without disorder in accordance with typical Lindemann numbers for crystals
\cite{GFCM2}. We note that this number does not
depend on the magnetic field which specifies the point on the melting line.
In the following we denote $ c_{44}(K_{\rm BZ}/\sqrt{2} ,0) $
and $ c_{66}(K_{\rm BZ}/\sqrt{2} ,0) $ in final expressions as for example in
(\ref{62}) by the abbreviations  $ c_{44}$ and $ c_{66} $.

From (\ref{55}) we can easily
calculate the BG-VG, VG-VL line.
By taking into account  $ c_L^2 a^2 \ll \xi'^2 $
which results in $ {\cal D}(2 \langle u^2 \rangle)
\approx {\cal D}(0) $ for YBCO we obtain for the unified
BG-VG, VG-VL line
\begin{equation}
  B_m  \approx
\frac{ \phi_0^5 \, (1-b)^{3} }{(k_BT)^2 \lambda^2_{ab}\lambda^2_c}
 \frac{ 3.9 \cdot 10^{-5} }{\pi^4} \; e^{- {\cal D}_0 (0)}  \label{65}
\end{equation}
Here we used $ a=\sqrt{\phi_0}/\sqrt{B} $ and
the typical defect length $ a_3 \approx  4 a \sqrt{2} \,\lambda_{ab}/
\lambda_c \sqrt{\pi} (1-b)^{1/2}$  \cite{Dietel2},
which results in the  disorder function
\begin{equation}
{\cal D}_0(0)  \approx   \left(d(T)\frac{ \xi_{ab}^2}{\xi'^2_0} \right)
 \frac{3.2\, (1-b)^{3/2}}{(k_B T)^2}
 \frac{\phi_0^4 \,
\xi_{ab}}{\lambda_{ab}^3 \lambda_{c}} \frac{\phi_0^{1/2}}{B^{1/2}}
\,.
\label{70}
\end{equation}
Note that (\ref{65}) is valid irrespective of the disorder correlation
function.

Parameter values for optimally doped
YBCO were given by Ref.~\onlinecite{Kamal1} as $ \lambda_{ab}(0) \approx 1186 $\AA
, $ \xi_{ab}(0)
\approx 15 $\AA,
$ T_c=92.7 K $.
The
 CuO$_2$ double layer spacing is
 $a_ s=12$\AA,
 and
the anisotropy parameter $ \gamma\equiv
\lambda_{c}/ \lambda_{ab} $
is approximately equal to  $ 5 $.
 From (\ref{70}) we obtain a unified BG-VG, BG-VL line which
scales like $ B_m \sim e^{-A_Y /\sqrt{B_m}} $.
Here $ A_Y $ is some constant independent of $ B_m $.
This results in a  BG-VG, BG-VL line
showing a reentrant behavior.
In Fig.~2 we show $ B_m $ with $ \delta T_c $-pinning on the upper
figure and $ \delta l $-pinning on the lower figure
for various values $ d_0 $. $ d_0 $ of the straight line curves
is chosen such that we have approximately
the best experimental curve fitting to the
 BG-VG, BG-VL curve of   Bouquet {\it et al.} \cite{Bouquet1}
shown by the dashed line with square points.
We obtain in fact a reentrant behavior of the
BG-VG, BG-VL   line. This is in accordance with the quadratic in disorder
calculation  for YBCO within the
Ginzburg-Landau approach by Li {\it et al.} \cite{Li1}. A reentrant behavior
of the BG-VG line was also seen in the experiments of
Pal {\it  et al.} \cite{Pal1} and Stamopoulos {\it  et al.}
\cite{Stamopoulos1}.
We must clarify that these experiments
are in contradiction to the majority of experiments which do
not see any reentrant behavior of the BG-VG line (see for example
\cite{Bouquet1,Deligiannis1,Radzyner1,Nishizaki1, Griessen1}).
The discrepancy in the shape of  the BG-VG transition line
lies presumably in different physical setting of these experiments
to the standard ones showing no reentrant behavior.
The experiment of Pal {\it  et al.} uses a crystal with a low density
of twins which could lead to deviations in the shape of the BG-VG line
due to \cite{Miu1}. The experiments of Stamopoulos {\it et al.} measures the
ac permeabilities which drives the crystal out of thermodynamical
equilibrium.

The solid lines with the stars at the left hand side
of Fig.~2 are calculated by solving (\ref{55}) restricted
to the transverse fluctuations with the
elastic moduli in the range $ b \lesssim 0.2 $ given in
\cite{Brandt1,Dietel2}. To calculate the transition curves
with the elastic moduli
$ b \lesssim 0.2 $  as well
is rather important because the solid curves in Fig.~2
calculated with moduli $ b \gtrsim 0.5 $
reaches immediately the range $ b \lesssim 0.2 $.
Note that the solid curves and the solid curves with stars are calculated
by using the same disorder constant $ d_0 $ of values
 $ 2\pi d_0 \, \xi_{ab}^2/ {\xi'}^2
=8.5 \cdot 10^{-8} $ ($ \delta T_c $- pinning)
and $ 2\pi d_0 \, \xi_{ab}^2/ {\xi'}^2
 =1.01 \cdot 10^{-6} $ ($\delta l$-pinning).

We obtain from Fig.~2 that the curves
of the $\delta l $-pinning mechanism
fits much better to the experimentally given  BG-VG, BG-VL line than
the $\delta T_c $-pinning curves.
This is in accordance to the observation in Ref.\,\onlinecite{Griessen1}.
Whatever the correct experimental BG-VG, BG-VL line shows a reentrant
behavior or not, it is not satisfactory within our approach
which is restricted to second order in the impurity potential,
that the solid curves with the stars ($ b \lesssim 0.2 $)
goes to zero at $ T \approx T_c $.
That this is true can be best seen in a logarithmic plot of the
BG-VG, BG-VL transition line $ B_m $ (\ref{65})
which is shown in Fig.~3. The straight line and the straight line
with the stars correspond to the curves of the $ \delta l $-pinning
mechanism shown in the lower part in Fig.~2.
The dashed curve with the triangle points is the BG-VG, BG-VL line
measured by Pal
{\it et al.} in Ref.\,\onlinecite{Pal1} mentioned above. This curve
 shows a
reentrant behavior. Both curves are in disagreement
at small magnetic fields.
Thus, we should go beyond second order in the disorder strength to get
better accordance with the experiments. This will be done in the following
sections.

\section{Replica variational method
of M\'ezard and Parisi}

In order to go beyond second-order perturbation theory in the impurity
potential, we use the well known replica trick \cite{Edwards1}
$ \ln Z = \lim_{n \to 0} (1/n)(\overline{Z^n}-1) $ where the overline means
disorder averaging and
\begin{align}
& \overline{Z^n}   =   \left[\prod_{\alpha}
\int {\cal D}
[u^\alpha_i,\sigma^\alpha_{im},n^\alpha_i]\right]   \nonumber \\&
~~~~~~~~~~~~~~~~~~
\times
 \overline{e^{
-\Sigma_{\alpha}
 \left(H_0[u^\alpha_i,\sigma^\alpha_{im},n^\alpha_i]+
H_{\rm dis}[u^\alpha_i]\right) /k_BT
}}          \nonumber \\
& =    \left[\prod_{\alpha}\int {\cal D}
[u^\alpha_i,\sigma^\alpha_{im},n^\alpha_i] \right]
e^{
-\Sigma_{\alpha}
 H_0[u^\alpha_i,\sigma^\alpha_{im},n^\alpha_i] /k_BT}
\nonumber \\
&
~~~~~~~~~~~~~~~~~~
\times
e^{- \Sigma_{\alpha,\beta}
H'_{\rm dis}[u^\alpha_i, u^\beta_i]/k_BT},   \label{85}
\end{align}
with\mn{looks like one $k_BT$ zuviel}
\begin{equation}
H'_{\rm dis}[u^\alpha_i,u^\beta_i]= \frac{-1}{2 k_B T}
\sum_{{\bf x}, {\bf x}'} \delta_{x_3, x_3'}
\Delta(x_i+u_i^{\alpha}({\bf x})-
x'_i-u_i^{\beta}({\bf x}')) \,.   \label{90}
\end{equation}
Here, the extra $ k_B T $ term in the denominator in (\ref{90}) comes from the
disorder average.

The average $ \overline{Z^n} $ cannot be calculated without
further approximations. In the following we use the low- and high-temperature
approximations of Section III for the result after the integration
over the stress fields $ \sigma_{ij}^\alpha $  and
defect fields $ n_{i}^\alpha $  in   $ \overline{Z^n} $ (\ref{85}).
Thus, we have to calculate partition functions of the following form
\begin{eqnarray}
\overline{Z^n}&  = & \left[\prod_{\alpha} {\cal N} \prod_{{\bf x},i}\left[
 \int_{-\infty}^\infty\frac{u^\alpha_i({\bf x})}{a} \right]\right] \nonumber \\&\times &
e^{-H/k_BT}
,   \label{100}
\end{eqnarray}
with the total Hamiltonian
\begin{eqnarray}
 H = \sum_\alpha H_0[u_i^\alpha]+\sum_{\alpha,\beta}
H'_{\rm dis}[u_i^\alpha, u_i^\beta] ,
\label{@HHH}\end{eqnarray}
where
$ H_0 $ is given by (\ref{19}) in the solid phase and by (\ref{25})
in the fluid phase.
Both are complicated
expressions which will need further approximations. The complications comes from
the large replica mixing interaction part $ H'_{\rm dis} $ in (\ref{85}).
In the following, we shall use a variational replica
method which was first given by
 M\'ezard and Parisi \cite{Mezard1}. With the help of this method
also used before for random spin models \cite{Dotsenko1} they
were able to calculate the glass transitions of isotropic
random manifold systems. These systems are described by the Hamiltonian
\begin{equation}
H_{\rm RM}= \int d^{d-N'}\!\! x  \, [ -{\bf u}({\bf x})
 ( {\bol \nabla} \cdot {\bol \nabla}) {\bf u}({\bf x})  +V({\bf u})] \,.
\label{110}
\end{equation}
Here $ {\bf u} $ is an $ N' $-dimensional vector describing an
$ N' $-dimensional manifold embedded in a $ d $-dimensional space.
$ V $ is an impurity potential with
a certain correlation function. When comparing the solid Hamiltonian
(\ref{19}) with the random manifold Hamiltonian (\ref{110}) and further by
setting the correlation length $ \xi'=0 $ in (\ref{36}) we obtain
that the transversal part of (\ref{19}) looks similar to a random manifold
with $ d-N'=3 $
in $ d=5 $ dimensions \cite{Mezard1} 
with a delta-like impurity correlation potential.
In the fluid phase described
by the high-temperature Hamiltonian (\ref{25}) we obtain for the transversal
part a random manifold with $ d-N'=1 $ and $ d=3 $ well known as a string 
embedded in three  dimensions.
The difference to the random manifold system comes then mainly
from the discretization in the third direction by the
dislocation length $ a_3 $ relevant
in the fluid phase as will be shown below.
It is well known that there exist for $ N' > 2 $
in a $ d=N'+1 $ random manifold system
corresponding to a string in $ d $ dimensions a roughening
transition separating
a low-temperature disorder dominated phase from a high-temperature
thermal phase \cite{Halpin1}. For $ N'<2 $ this phase transition
is not existent and the system is dominated mainly by disorder 
fluctuations. It is now believed by computer simulations
that at the critical dimension $ N'=2 $ corresponding
to a string in three dimensions with a $ \delta $-correlated impurity
potential the roughening transition
of the string system  is described by a
crossover \cite{Halpin1}. Below we show that this roughening
transition corresponds to the glass transition of the vortex lattice.
That  the vortex lattice at $ d=3 $ is in fact
at the lower critical dimension for a glass transition was mentioned before
for an XY-model of the gauge class type \cite{Reger1}. This XY-model
as similar ones with other disorder potentials mentioned in the introduction
are toy model for a disordered vortex lattice in superconductors.

The M\'ezard-Parisi theory   consists in replacing
the non-quadratic part of the Hamiltonian
as quadratic with a possible  mixing
of replica fields. By using the Bogoliubov variational principle we can find
the best matrix of this quadratic form so that the free energy of the
variational Hamiltonian is as close as possible to the actual free energy
of the system. This means that we have to search the minimum of the
variational free energy
\begin{equation}
 F_{\rm var}= F_{\rm trial} +\langle H- H_{\rm trial} \rangle_{\rm trial}    \label{115}
\end{equation}
with the harmonic trial Hamiltonian
\begin{equation}
H_{\rm trial} = \frac{v}{2}\sum_{{\bf x},{\bf x}'}
\sum_{\alpha, \beta}  {\bf u}^{\alpha}({\bf x})
{\bf G}^{-1}_{\alpha \beta}
({\bf x}-{\bf x}')
{\bf u}^{\beta}({\bf x}') \,.  \label{120}
\end{equation}
Here $ \langle \cdot \rangle_{\rm trial} $ stands for the averaging with
respect of the Gibb's measure of the trial Hamiltonian
$ H_{\rm trial} $, and $ F_{\rm trial} $ denotes
 the associated free energy.
In Section~ IX,  we shall use the intersection criterion
with $   F_{\rm var} $ for the solid and fluid phase to determine the
BG-VG, BG-VL transition line. By using (\ref{120}), we obtain
the free energy associated with
(\ref{100}):
\begin{align}
& F_{\rm var}=-k_B T \frac{N }{2 V_{\rm BZ}}\int_{\rm BZ}
 d^2 k dk_3 \; \bigg(\ln\left[{\rm det} \,
\left( \frac{2 \pi k_B T}{v a^2} {\bf G}\right) \right]
\nonumber \\
& + {\rm Tr} \left\{\left[{\bf G}^{-1}({\bf k})-
{\bf G}_0^{-1}({\bf k})\, {\rm I}
\right] {\bf G}( {\bf k}) \right\}  \bigg)
\nonumber \\
&  -k_B T \ln({\cal N})  +
\left\langle \sum_{\alpha, \beta} H'_{\rm dis}[u_i^\alpha, u_i^\beta]
\right\rangle_{\rm trial}     \,.   \label{125}
\end{align}
where we use bold symbols for 
vectors
and matrices in the vortex displacement plane.
The symbol ${\rm I} $
denotes the unit matrix in replica space.
The trace
$ {\rm Tr}[\,\dots\,] $
 runs over the replica  indices and
vortex displacement indices.
In principle we can obtain a general expression
for the disorder term given by
the last term in (\ref{125}). Because one should use different
approximations for the solid phase and the liquid phase, we shall
give directly approximations for this term in both phases
at the beginning of the following sections.

It will be clear soon for the solid as well as the fluid phase
that $ {\bf G}_{\alpha \beta} $ can be chosen to have the form
\begin{equation}
{\bf G}^{-1}_{\alpha \beta}=
{\bf G}_0^{-1} \delta_{\alpha \beta }+ \sigma_{\alpha \beta}\, {\bf I}      \label{130}
\end{equation}
where $ {\bf I}  $ is the two-dimensional unit matrix in
the vortex displacement plane. To find a local
minimum of (\ref{125}) in the space of all symmetric self-energy matrices
$ \sigma_{\alpha \beta} $ was simplified considerably by Parisi in the
case of spin glasses. There he restricted the search of the minimum
for (\ref{115})
to the case of some sort of closed algebra known as the algebra of
Parisi matrices \cite{Dotsenko1, Mezard2}.
In Appendix~B we prove some stability theorems for stationary points
of $ F_{\rm var} $ (\ref{125}).
These are summarized at the end of Appendix~B2. The restriction
of the minimum search to self-energy matrices in the Parisi-algebra
is justified among others by the fact
that a local minimum within the Parisi-algebra
is automatically  a local minimum in the whole self-energy space
without the restriction to the Parisi-algebra.
This is shown in Appendix~B.

In general the minimum
self-energy matrix $ \sigma_{\alpha \beta} $ is not
symmetric under the interchange of replica indices which means
that the local minimum $  \sigma_{\alpha \beta} $
of $ F_{\rm var} $ (\ref{125})
is not unique. This is typical for glasses where
the minimum of the free energy is degenerate \cite{Dotsenko1}.
This degeneracy corresponds to the degeneracy of the stable states in
glasses with high energy barriers between them. These are
responsible for the irreversibility phenomena
beyond the glass transition lines in high-temperature superconductors
mentioned in the introduction.

\section{Fluid  phase}
In this section, we derive  the variational free energy $ F_{\rm var} $ in the
liquid phase. We obtain
\begin{align}
&\left \langle \sum_{\alpha, \beta} H'_{\rm dis}[u_i^\alpha, u_i^\beta]
\right\rangle_{\rm trial} \approx -  \frac{N}{2 k_B T} \sum_{\alpha,\beta}
 \frac{1}{(2 \pi)^2}         \nonumber \\
& \times  \int d^2 q \;  \hat{\Delta}(q)    e^{- \frac{1}{2}{\bf q} \! \cdot  \!
\bigg[ \! {\bf G}_{\alpha \alpha}(0)+
{\bf G}_{\beta \beta}(0)  -
{\bf G}_{\alpha\beta}(0)-
{\bf G}_{\beta \alpha}(0) \!  \bigg]{\bf q}} \nonumber \\
& \approx -  k_B T \, \frac{N}{2}
 \sum_{\alpha, \beta}   {\cal D}\left(2 B_{\alpha \beta} \right)  \label{140}
\end{align}
with
\begin{equation}
B_{\alpha \beta} = \frac{k_B T}{2 v}{\rm tr}
\left[{\bf G}_{\alpha \alpha}(0)
+{\bf G}_{\beta \beta}(0)  -
{\bf G}_{\alpha\beta}(0) -
{\bf G}_{\beta \alpha}(0) \right]   \,.      \label{150}
\end{equation}
where $ \tilde{\Delta} ({\bf q}) $
is the two-dimensional
Fourier transform of $ \Delta({\bf x}) $.
{The trace
$ {\rm tr}[\,\dots\,] $
 runs over the vortex displacement indices.}
In (\ref{140}) we restricted us
in the double sum over $ {\bf x} $, $ {\bf x}' $ on the diagonal summands
$ {\bf x} = {\bf x}' $. The reason for the validity of this restriction
comes from the observation that due to (\ref{25}) the non-diagonal
summands are given by
\begin{align}
&  \left[\left\langle \sum_{\alpha, \beta} H'_{\rm dis}[u_i^\alpha, u_i^\beta]
\right\rangle_{\rm trial}\right]_{{\bf x}\not={\bf x}'}\!\!\!\!=
  -\frac{1}{2 k_B T}
\sum_{{\bf x} \not= {\bf x}'}
\nonumber \\
& \times  \delta_{x_3, x_3'} \sum_{\alpha,\beta} \frac{1}{(2 \pi)^2} \int d^2 q \;  \hat{\Delta}({\bf q})
 e^{i q_i (x_i-x_i') }\label{160}  \\
&
\times  e^{  - \frac{1}{2}{\bf q} \cdot \left[{\bf G}^T_{\alpha \alpha}(0)+
 {\bf G}^T_{\beta \beta}(0)
 -{\bf G}^T_{\alpha\beta}({\bf x}-{\bf x}')-
{\bf G}^T_{\beta \alpha}({\bf x}-{\bf x}')
\right]
{\bf q}}
        \nonumber \\
&
\times  e^{- \frac{1}{2}{\bf q}
\cdot \left[{\bf G}^L_{\alpha \alpha}(0)
 + {\bf G}^L_{\beta \beta}(0) -
{\bf G}^L_{\alpha\beta}({\bf x}-{\bf x}')-
{\bf G}^L_{\beta \alpha}({\bf x}-{\bf x}')\right] {\bf q}}
\nonumber
\end{align}
where $ {\bf G }^L = {\bf P}_L {\bf G } {\bf P}_L $ and
$ {\bf G }^T = {\bf P}_T {\bf G } {\bf P}_T
$ are the longitudinal and transversal
components of the Green function. By using (\ref{25})
the second exponent in (\ref{160})
corresponding to the transversal part of the vortex fluctuations
can be tranformed to
\begin{align}
&  - \frac{1}{2}\bigg\{ ({\bf q}^T)^2(0)  \,
 [G^T_{\alpha \alpha}(0)+
   G^T_{\beta \beta}(0)]
   \nonumber  \\
  &
 -
 ({\bf q}^T)^2
 ({\bf x}-{\bf x}')\;
 [G^T_{\alpha\beta}(0)+
G^T_{\beta \alpha}(0)]
\bigg\}            \label{163}
\end{align}
where $ G^{L,T}_{\alpha \beta}({\bf x}) =
{\rm Tr}[{\bf G}^{L,T}_{\alpha \beta}({\bf x})] $.
Due to the large thermal
effective one-dimensional transverse fluctuations we have either
$ G^T_{\alpha \beta}(0) \rightarrow  \infty $ where
$ G^T_{\alpha \alpha}(0)-G^T_{\alpha \beta}(0) $ is finite and
$ \alpha $, $ \beta $ is arbitrary or
$ G^T_{\alpha \alpha}(0) \rightarrow  \infty $ and $ G^T_{\alpha \beta}(0)$
is finite for $\alpha \not= \beta $
where in both cases  the self-energy matrix
is restricted to the Parisi algebra.
This is shown in Ref.~\onlinecite{Mezard1}.
From this we obtain the vanishing of (\ref{160}).

First, we take the variation of the free energy (\ref{125}) with
respect to 
the diagonal Green function matrix elements $ G_{\alpha \alpha} $.
This results in
\begin{equation}
\sum_\beta \sigma_{\alpha  \beta} = 0   \label{165}
\end{equation}
That the minimum of $ F_{\rm var} $ should be found in the symmetric
self-energy matrices with the constraint (\ref{165}) is suggestive because
(\ref{165}) justifies that the Hamiltonian (\ref{120}) has the global
translational symmetry $ {\bf u}^\alpha({\bf x}) \rightarrow
{\bf u}^\alpha({\bf x})+{\bf t} $ for any vector $ t $, which has also the
disorder Hamiltonian  (\ref{90}).

In the most general case within the Parisi-algebra, the form
of the self-energy $ \sigma_{\alpha \beta} $
with the constraint (\ref{165}) can be
described by a continuous function $ \sigma(s) $ with $ 0 < s < 1 $
\cite{Mezard1}. In that case the trial free energy takes the form
\begin{align}
& \Delta f_{\rm var}  \equiv  \frac{1}{N} \lim_{n \to 0} \frac{1}{n}\left[
F_{\rm var}(B[\Delta]) - F_{\rm var}(0) \right]     \label{170}     \\
 & =    \frac{k_B T }{2 } \int_0^1 ds \; \bigg[ \frac{1}{s^2}
\int_0^{\Delta(s)} \! \! \! \! \!
d \Delta  \Delta \frac{d}{d \Delta} g(\Delta)  +
{\cal D}_0 \left(2 B[\Delta(s)] \right) \bigg]   \nonumber \\
& f_{\rm var}(0)  =    \frac{1}{N}
{\rm lim}_{n \to 0} \frac{1}{n} F_{\rm var}(0)=-k_B T \bigg( \frac{1}{N}
\ln {\cal N}
\label{175}   \\
&  + \frac{1}{2} \bigg\{\frac{1}{
 V_{\rm BZ}} \int_{\rm BZ}
 d^2 k dk_3 \; \ln \left[{\rm det} \, \left(
\frac{2 \pi k_B T}{v a^2}  {\bf G}_0 \right)\right]
+   {\cal D}_0(0) \bigg\}\bigg)
\nonumber
\end{align}
 where
\begin{equation}
g(\Delta) =  \frac{1}{ V_{\rm BZ}} \int_{\rm BZ}  d^2k dk_3
{\rm Tr} \left[\left({\bf G}_0^{-1} +
\Delta  {\bf I} \right)^{-1} \right]  \,.   \label{180}
\end{equation}
The gap function $ \Delta(s) $ and the self-energy function
$ \sigma(s) $ corresponding to the self-energy matrix
$ \sigma_{\alpha \beta} $ in the non-continuous case is related by
\begin{equation}
\Delta(s) = \int_0^s  ds' s' \frac{d \sigma(s')}{d s'} \,.\label{190}
\end{equation}
$ B[\Delta(m)] $ corresponding to $ B_{\alpha \beta} $ (\ref{150})
in the continuous
case is given by
\begin{equation}
B[\Delta(s)]=  \frac{ k_B T}{v} \frac{1}{s} g[\Delta(s)] -
\frac{k_B T}{v}
\int_s^1 ds'
\frac{1}{s'^2} g [\Delta(s')]  \,.                \label{200}
\end{equation}
In order to find the local minimum of $ f_{\rm var} $ we have
to take the derivative of (\ref{170}) with respect to $ \Delta(m) $.
This results in
\begin{equation}
\sigma(s) = -  2 \,  \frac{k_B T}{v} \;
{\cal D}'_0 \left(2 B[\Delta(s)] \right)  \,,
\label{210}
\end{equation}
where $ D'(x) $ is the derivative $ (d/dx) D(x) $.
We point out that (\ref{210}) shows that
\begin{equation}
 \sigma(s) \ge 0  \quad , \qquad   \Delta(s)\ge  0 \,.   \label{212}
\end{equation}
In the following, we discuss solutions of  this equation in the
case that $ \sigma(s) $ does not break the replica symmetry,
is one-step replica symmetry-breaking or continuous replica symmetry-breaking.
\subsection{Symmetric solution}
We now solve Eq. (\ref{210}) for $ \sigma(s) $ with an Ansatz which does not
break any replica symmetry.
The Ansatz for $ \sigma(s) $ in this case is
\begin{equation}
\sigma(s) = \sigma_0   \,.                \label{215}
\end{equation}
By using the equations (\ref{190}) and (\ref{210}) we obtain
\begin{eqnarray}
\sigma(s) & = &   0   \,,                           \label{217}   \\
 \Delta(s) & = &  0   \,.                                  \label{220}
\end{eqnarray}
From this we obtain that $ B[\Delta(m)] \rightarrow \infty $
for infinite area of the system. This results in
\begin{equation}
\Delta f_{\rm var} = 0   \,.         \label{230}
\end{equation}

\subsection{One-Step replica symmetry-breaking}
The simplest possible extension of the replica symmetric case
above consists of a one-step replica symmetric solution given by
\begin{equation}
\sigma(s) = \left\{ \begin{array}{ccc}
             \sigma_0   &  \mbox{for} & 0  < s <m_1  \,,  \\
             \sigma_1   & \mbox{for} & m_1 < s <1 \,.
              \end{array} \right.   \label{235}
\end{equation}
By using this Ansatz in (\ref{170}) we obtain
\begin{align}
& \Delta f_{\rm var} =  - \frac{k_B T}{4} \left(1-\frac{1}{m_1}\right)
\nonumber \\
&
\times \left[
\left(\frac{\tilde{\Delta}_1}{1+ \tilde{\Delta}_1/4}\right)^{1/2}
\!\!\!\!\! \! \! \!\! \!-
4\, {\rm arcsinh}\left(\frac{\tilde{\Delta}^{1/2}_1}{2}\right)\right]
\nonumber \\
&  + \frac{k_B T }{2}
(1-m_1)
{\cal D}
\left( 2 \frac{k_B T }{v} g(\Delta_1) \right)
\label{240}
\end{align}
where we used $ \sigma_0 =0 $ which can be derived
from (\ref{210}) and (\ref{190}) similar to the replica symmetric case.
Furthermore, we used the abbreviation $ \Delta_1 =m_1(\sigma_1-\sigma_0) =
m_1 \sigma_1  $ and $ \tilde{\Delta}_i \equiv  \Delta_i a_3^2
/c_{44} $.
We restricted us in the calculation of $ \Delta f_{\rm var} $ to the
transversal components of $ {\bf G_{0}} $  which effectively means
that we set $ G_{0}^L=0  $ in the calculation $ g(\Delta) $ in
(\ref{180}). The longitudinal term in $ g(\Delta) $ is a factor
$ a^2 c_{44}/a^2_3 c_{11} = (a^2 c_{44}/a^2_3 c_{66}) (c_{66}/c_{11})
\approx (\pi /4) (c_{66}/c_{11}) $ smaller which is
 justified by $ c_{66} \ll c_{11} $
\cite{Brandt1} irrespective of the value of $ \tilde{\Delta}_1 $.
For the derivation of $ \Delta f_{\rm var} $ we used for $g(\Delta)$ in
(\ref{180})
\begin{eqnarray}
 g(\Delta) & \approx & \frac{a_3^2}{c_{44}}  \frac{1}{V_{\rm BZ}}
 \int_{\rm BZ}  d^2k dk_3
 \frac{1}{[2-2 \cos(k_3 a_3)]+ \tilde{\Delta}} \nonumber \\
 & = &   \frac{1}{2} \frac{1}{ \tilde{\Delta}^{1/2}(1+\tilde{\Delta}/4)^{1/2}}
 \frac{a^2_3}{c_{44}} \,.   \label{250}
\end{eqnarray}
We now determine the stationary point of $ \Delta f_{\rm var} $ (\ref{240}).
By setting the derivative of $ \Delta f_{\rm var} $
with respect to $ \Delta_1 $
and $ m_1 $ equal to zero we obtain two equations for the stationary values
of $ \Delta_1 $ and $ m_1 $.
These are given by
\begin{align}
&   \frac{1}{8} \left(1-\frac{1}{m_1}\right)\tilde{\Delta}_1^{-1/2}
= \frac{1}{4}\left(1-m_1\right)
\frac{k_B T }{ \tilde{\Delta}_1^{3/2}} \frac{a_3}{c_{44} a^2} \nonumber \\
& \times {\cal D}'\left( 2 \frac{k_B T }{v} g(\Delta_1) \right)
  \,,
\label{253} \\
& \frac{1}{4}\frac{1}{m_1^2}
\left[ 4 {\rm arcsinh}\left(\frac{\tilde{\Delta}^{1/2}_1}{2}\right)-
\left(\frac{\tilde{\Delta}_1}{1+ \tilde{\Delta}_1/4}\right)^{1/2}
\right]  \nonumber \\
&   =\frac{1}{2} \; {\cal D}
\left( 2 \frac{k_B T }{v} g(\Delta_1) \right)  \,.
\label{255}
 \end{align}
In the following solution of (\ref{253}) and (\ref{255}) we use that
$ \tilde{\Delta}_1 \ll 1 $ in the interesting range near the glass transition
line which we expect at $ \tilde{\Delta}_1=0 $. This will be shown below.
These two equations can be solved exactly in this limit resulting in
\begin{eqnarray}
m_1^3 & = & ({\cal D}(0) A)^{-1} \,, \label{260} \\
\tilde{\Delta}_1^{1/2} & = & 2 A^{-1}
\left(\frac{1}{m_1}-1 \right)
\label{270}
\end{eqnarray}
where constant $ A $ similar to the Lindemann constant written
for general disorder correlation functions
(see the definitions (\ref{33})-(\ref{38}))
\begin{equation}
A_{0, \infty}=
\frac{4}{k_B T} \frac{ c_{44} a^2 \; {\xi'}^2_{0, \infty} }
{a_3} \,.   \label{275}
\end{equation}
with $ A \equiv A_{0}=   A_{\infty} $ for the Gaussian correlation function
(\ref{15}).
Here, we mention that $ A \approx b /2\pi c_L^2 \gg 1 $ near the melting line
without disorder $ V= 0 $ \cite{Dietel2}. This is the
magnetic-temperature regime, we are interested in.
By the help of (\ref{240}), (\ref{260}) and (\ref{270})
we can calculate the free energy $ \Delta f_{\rm var } $ getting
\begin{equation}
\Delta f_{\rm var}=\frac{k_B T }{2} {\cal D}_{\infty}(0) \left[1-
({\cal D}_{\infty}(0) A_{\infty })^{-1/3}  \right]^3
\label{280}
\end{equation}
for $ {\cal D}_{\infty}(0) A_{\infty} \ge 1 $ in the regime
$ (({\cal D}(0) A)^{1/3}-1)/A  \ll 1 $ for the Gaussian correlation function
(\ref{15}).
As suggested by the indices, expression (\ref{280}) is more general
valid irrespective
of the disorder
correlation potential  in the restricted regime
$ (({\cal D}_{\infty}(0) A)_{\infty}^{1/3}-1)  \ll 1 $ (see the discussion above Eq.~
(\ref{920})).
Next, we must calculate also the replica symmetry-breaking solutions
of the free energy (\ref{170}) having more than one discrete step. To solve
the minimum problem in this case is rather difficult. Therefore, we restrict
us first to the determination of the continuous symmetry-breaking solutions.

\subsection{Continuous symmetry breaking}
Finally, we look for solutions $ \sigma(s) $ of (\ref{210}) which are
continuous. In this case, we can solve the stationary equation by a partial
integration of $ B[\Delta(s)] $ in (\ref{200}) resulting in \cite{Giamarchi1}
\begin{equation}
B[\Delta(s)]= B[\Delta(s_c)]-\frac{k_B T}{v}
 \int_s^{s_c} ds' \sigma'(s')
g'(\Delta[s'])  \,.  \label{500}
\end{equation}
Here we assumed that $ \sigma(s) = {\rm constant} $ for $ s \ge s_c $.
By taking two derivatives of (\ref{210}) we obtain that $ \sigma(s) $ fulfills
the following equation
\begin{equation}
\sigma'(s) = - 2 \sigma'(s) \sigma(s)^{3/2}
\frac{(k_B T)^{1/2}}{\xi' {\cal D}^{1/2}(0)} \frac{1}{v^{1/2}} g'(\Delta[s]) \,. \label{510}
\end{equation}
Similar as in the case of the one-step symmetry-breaking solution
we can neglect the longitudinal component in $ g(\Delta) $ (\ref{180})
being a factor $ c_{66}/c_{11} \ll 1 $ smaller than the transverse term in
$ g(\Delta) $ (see the discussion below (\ref{240})).
We point out that this is true irrespective of the value of
$ \Delta $. Using (\ref{250}) we
obtain two solutions of (\ref{510}) by taking once more the derivative
with respect to $ s $. This results in the following solutions of (\ref{510})
\begin{align}
& 1. \quad \sigma'(s)=0 \,,   \label{520} \\
& 2. \quad
\left(\frac{a^2_3}{c_{44}}\right)  \sigma(s) \, s
=\frac{\tilde{\Delta}(s) \left[1+ \frac{1}{4}\tilde{\Delta}(s)\right]
\left[1+ \frac{5}{4}\tilde{\Delta}(s)\right]}{1+
\frac{2}{3}\tilde{\Delta}(s) +
\frac{1}{6}\tilde{\Delta}^2(s) } \,.  \label{530}
\end{align}
By inserting (\ref{510}) into  (\ref{530}) we obtain
for the second type of solutions
\begin{equation}
  s \,({\cal D}(0) A)^{1/3} = \frac{\left[1+ \frac{5}{4}
\tilde{\Delta}(s)\right]^{5/3}}
{1+ \frac{2}{3}\tilde{\Delta}(s) + \frac{1}{6}\tilde{\Delta}^2(s)}
\,.    \label{535}
\end{equation}
Finally, we determine the constant $ s_c $ defined in (\ref{500})
where $ \sigma(s) = {\rm constant} $ for $ s_c < s < 1 $.
By using (\ref{210}) we obtain
\begin{equation}
\sigma(s_c) = - 2 \frac{k_B T}{v}   \,
{\cal D}' \left(2 B[\Delta(s_c)] \right)  \,.
\label{540}
\end{equation}
With the help of (\ref{530}) we obtain
\begin{eqnarray}
  {\cal D}(0) A s_c & = & \frac{\left[1+ \frac{5}{4}\tilde{\Delta}(s_c)\right]
}{1+
\frac{2}{3}\tilde{\Delta}(s_c) +
\frac{1}{6}\tilde{\Delta}^2(s_c) }   \label{550}   \\
 & & \times  \left\{\frac{1}{2} A \;\tilde{\Delta}^{1/2}(s_c)
\left[1+  \frac{1}{4}\tilde{\Delta}(s_c)\right]^{1/2}+1\right\}^2
  \nonumber
\end{eqnarray}
which leads with (\ref{535}) to
\begin{equation}
({\cal D}(0)A)  = \frac{\left\{\frac{1}{2} A \;\tilde{\Delta}^{1/2}(s_c)
\left[1+  \frac{1}{4}\tilde{\Delta}(s_c)\right]^{1/2}+1\right\}^3}{\left[1+ \frac{5}{4}\tilde{\Delta}(s_c)\right]}
\,.
\label{560}
\end{equation}
Under consideration of  (\ref{212}) we obtain that (\ref{560}) can be
solved only for $ {\cal D}(0) A \ge 1 $.
Furthermore, by taking into account (\ref{535})
in the case of $ \tilde{\Delta}= 0 $ which is the same equation
when taking only
the most leading $ \tilde{\Delta} $-term for $ \tilde{\Delta} \rightarrow 0 $
in (\ref{250}), we obtain in this limit no solution of (\ref{535}).
As mentioned above, this corresponds to the marginality of a string in
$ d=3 $ dimensions in an impurity
background.
Due to the non-quadratic polynomial behavior
of the expressions above, it is not possible to get simple
analytic solutions in the whole $ \tilde{\Delta}$-range.
Therefore, we shall solve (\ref{520}),
(\ref{535}) and  (\ref{550})  for small $ \tilde{\Delta} \ll 1 $.
This corresponds to the restriction $ (({\cal D}(0) A)^{1/3}-1)/A  \ll 1 $.
We obtain in this range the following solutions:
\begin{equation}
\tilde{\Delta}(s)  = \!\left\{ \!\!
\begin{array}{ccc}
0 & \mbox{for} & 0 \le s \le  \frac{1}{ ({\cal D}(0)A)^{1/3}}  \\
\frac{12}{17} \left[ ({\cal D}(0)A)^{1/3} s -1\right]  & \mbox{for}
&    \frac{1}{  ({\cal D}(0)A)^{1/3}}  \le s \le s_c \\
\frac{4}{A^{2}} \left[({\cal D}(0) A)^{1/3} -1\right]^2 & \mbox{for} &
 s_c \le s \le 1 .
\end{array} \right. \label{570}
\end{equation}
with
\begin{equation}
s_c  \approx   \frac{1}{ ( {\cal D}(0) A)^{1/3}} +\frac{17}{3} {\cal D}^2(0)
\frac{(({\cal D}(0) A)^{1/3}-1)^2}{({\cal D}(0) A)^{7/3}}
\label{575}
\end{equation}
Finally, we can calculate the free energy $ \Delta f_{\rm var} $ for
the replica symmetry-breaking solution (\ref{570}) by
using (\ref{170}), (\ref{250})
 and (\ref{530}) for $  (({\cal D}(0) A)^{1/3}-1)/A  \ll 1 $. We obtain
\begin{equation}
\Delta f_{\rm var}=\frac{k_B T }{2} {\cal D}_{\infty}(0) \left[1-
({\cal D}_{\infty}(0) A_{\infty })^{-1/3}  \right]^3
\label{590}
\end{equation}
identical with the free energy of the one-step replica symmetry breaking 
solution (\ref{280}). 
(\ref{590}) is valid for $ (({\cal D}(0) A)^{1/3}-1)/A \ll 1 $ in the case
of a Gaussian disorder potential. But one can generalize the calculation above
to obtain the validity of (\ref{590})  in the smaller regime
$ (({\cal D}_\infty(0) A_\infty)^{1/3}-1) \ll 1 $
irrespective of the disorder
correlation function (see the discussion above Eq.~(\ref{920})).
Summarizing, we obtain a saddle point of
$ \Delta f_{\rm var} $ which is
symmetric in replica space for ${\cal D}_\infty(0)A_\infty  \le 1 $.
In the case of
${\cal D}_\infty(0)A_\infty  \ge  1 $
we obtain a replica-symmetric solution, a one-step
replica symmetry-breaking solution and also a continuous
replica symmetry-breaking solution appears. To get more insight into
the true minimum, we have to consider the stability of the various
saddle point solutions in this case.

\section{Stability of solutions}
In this section we determine whether the various solutions for the fluid phase
discussed in the last section are stable in a sense specified below and
whether we have to take into account also higher-step replica
symmetry-breaking solutions to get a stable saddle point.
A typical example of an exactly solvable
system with finite-step replica symmetry-breaking
saddle point solutions which are not stable is a string in two dimensions
with a $\delta$- impurity correlation function
resulting in an unphysical
negative variance of the free energy with respect to disorder averaging
\cite{Saakian1}. This negative variance is vanished
in  the case of the infinite or continuous replica
symmetry-breaking solution.
M\'ezard and
Parisi \cite{Mezard1} show two different ways to obtain
a theory which includes replica symmetry breaking such as
(\ref{170})-(\ref{210}) for random manifolds.
The first approach consists of a large $ N' $
expansion of the partition function
where $ N' $ is the number of components of the fields.
There is only a slight difference between the large $ N' $ approach
and the variational approach used above. In the saddle point equation
of the large $ N'$-approach we have to substitute
$ {\cal D}(x) $ in (\ref{170})-(\ref{210}) by $ \Delta(\sqrt{x})/ (k_B T)^2 $
for the application of this approach to the fluid phase of the vortex
lattice. This is discussed in Appendix~B.
The large  $ N' $ expansion consists effectively in a saddle point
approximation in suitable chosen auxiliary fields \cite{Mezard1}.
The stability of solutions of these equations consists in going one
step further to the quadratic expansion of the
action in these auxiliary fields with the requirement that the
partition function calculated from this saddle point approximation
is not divergent when integrating out the auxiliary fields.
It was shown by Carlucci {\it et al.} \cite{Carlucci1} that continuous replica
symmetry-breaking  solutions calculated in the last subsection are generally
stable in this sense.
This is reviewed by us in Appendix~B1.
Due to the smallness of  $ N'=2 $ in the vortex
lattice system we do not think  that the large $ N'$-expansion
is appropriate in our case.

We derived (\ref{170})-(\ref{210}) by another way also stated first by
M\'ezard and Parisi \cite{Mezard1} via the variational approach in
(\ref{115}). It is clear that in this case we should require for the
eigenvalues of the matrix built of the
second derivatives of $ \Delta F_{\rm var} $ with respect to the self-energies
$ \sigma_{\alpha \beta} $ that these are all positive in the stationary point.
Here we take further into account the symmetry of $ \sigma_{\alpha \beta} $ and
(\ref{165}) in the variation of the free energy. The concrete
derivation  was carried out by \v{S}\'{a}\v{s}ik in Ref.~\onlinecite{Sasik1}.
Starting from his expression for the Hessian
we carry out in Appendix~B2 a similar stability analysis as was done
in the large $ N' $ case by Carlucci {\it et al.} in 
Ref.~\onlinecite{Carlucci1} summarized in Appendix~B1.
We also find in the variational approach
that the continuous symmetry-breaking solutions are generally
stable which means
that all eigenvalues of the Hessian are larger than or equal to zero.
Furthermore, we show in Appendix~B2 that the full Hessian has positive
or zero eigenvalues if and only if the replicon sector consists of
positive or zero eigenvalues. Thus it is enough to consider
only the replicon sector for stability.
The lowest eigenvalues in the replicon sector \cite{Carlucci1}
are given in (\ref{a210}) where $ k=l=r+1 $. Here
$ \tilde{f} $ is replaced by
the disorder function
$ {\cal D} $ in our case and $ L'_{kl} $ is given in
(\ref{a45}). $ G_0 $ is the transversal Green
with zero self-energy and
$ G_\alpha $, $ \Delta_\alpha $ are the value of the transversal
full Green function and gap function in the Paris block
$ 1 \le \alpha \le R+1 $ in a Parisi hierarchy of level $ R $.

We now come to a discussion of the stability of the saddle point
solutions in the fluid high-temperature phase
in the symmetric form given in Section~VIA and  the one-step
replica symmetric form given in Section~VIB.

First, we consider the saddle point solution given in Section~VIA
for the replica  {\it symmetric} form.
The most relevant replicon eigenvalue $ \tilde{\lambda}(0;1,1) $ is
given by (\ref{a210})
 \begin{equation}
\tilde{\lambda}(0;1,1) \propto 1 + 4  \frac{(k_B T)^2}{v^2}
D''\left(2 \frac{k_B T}{v} g(0)\right) g'(0)   \label{800}
\end{equation}
Here the proportionality factor is positive.
The stability criterion $ \tilde{\lambda}(0;1,1)\ge 0 $ leads to
\begin{equation}
{\cal D}(0) A \le 1 \,.   \label{805}
\end{equation}
Note that $ {\cal D}(0) A = 1 $ corresponds to $ m_1=1 $
in the one-step replica symmetric solution (\ref{260}).

Next, we consider the stability criterion for solutions of the
stationarity condition (\ref{210}) in the {\it one-step replica symmetry 
breaking} form. Here, we obtain the lowest replicon eigenvalues from
(\ref{a210})
 \begin{align}
& \tilde{\lambda}(0;1,1)  \propto  1  + 4  \frac{(k_B T)^2}{v^2}
  \label{810}  \\
&  \times
D''\left(2 \frac{k_B T}{v}
 \left\{g(\Delta_1)+\frac{1}{m_1} \left[g(0)-g(\Delta_1)
\right]\right\}\right)
 g'(0)
  \nonumber        \\
& \tilde{\lambda}(1;2,2)  \propto 1
 + 4  \frac{(k_B T)^2}{v^2} D''\left(2 \frac{k_B T}{v}
 g(\Delta_1)\right) \;
g'(\Delta_1) \,.  \label{820}
\end{align}
Because $ g(0) $ is divergent (\ref{250}) we obtain for the stability
criterion $ \tilde{\lambda}(0;1,1)\ge  0 $
\begin{equation}
m_1 \le 1   \,.    \label{830}
\end{equation}
By using (\ref{250}), (\ref{260}) and (\ref{270}) we obtain
$ \lambda(1;2,2)=0 $ in the leading order in
 $ \tilde{\Delta}_1 \ll 1, 1/A^2  $.
This can be seen much easier without using
the solutions (\ref{260}) and (\ref{270}) from equations (\ref{253})
(\ref{255}). By taking the square of Eq. (\ref{253}) times the inverse
of Eq. (\ref{255}) we obtain under the  consideration  of
$ D''(x)=2 D'(x)^2/D(x) $ which we obtain from
(\ref{36}) that $ \lambda(1;2,2)=0 $ when using
(\ref{255}) in the leading order in $ \tilde{\Delta}_1 $.
This now gives the possibility to calculate the stability
criterion also in the non-leading order in $ \tilde{\Delta}_1 $.
We obtain
 \begin{align}
& 1= 2\frac{(k_B T)^2 a_3^2 }{c_{44}^2 a^4 }\frac{1}{ \tilde{\Delta}_1^2}
\left[\!\!4 {\rm arcsinh}\left(\frac{\tilde{\Delta}_1^{1/2}}{2}\right)
\! - \! \left(\frac{\tilde{\Delta}_1}{1+ \tilde{\Delta}_1/4}\! \right)^{1/2}
\! \right]  \nonumber \\
& \qquad \times \frac{(D')^2 (2 k_B T
 g(\Delta_1)/v        ) }{D(2 k_B T
 g(\Delta_1)/v )}
\label{840}
 \end{align}
which results in
\begin{align}
& \tilde{\lambda}(1;2,2) \propto 1-\frac{(k_B T)^2 a_3^2 }{c_{44}^2 a^4 }
\frac{(1+5 \tilde{\Delta}_1/4)}
{\{\tilde{\Delta}_1[1+\tilde{\Delta}_1/4]\}^{3/2}}
\nonumber \\
& \times D''\left(
2 \frac{k_B T}{v}
 g(\Delta_1) \right) <0           \,.
\label{850}
 \end{align}
when taking the correlation function (\ref{36}) into account.
Summarizing, the one-step replica symmetry-breaking solution for
the correlation function (\ref{36}) is unstable.
Nevertheless, this instability is very weak for $ \tilde{\Delta} \ll 1 $
which is the interesting region near the glass transition line.
More generally, we show in Appendix~C that all  finite-step replica
symmetry broken
 solutions of the saddle point equation (\ref{210}) are unstable
for the Gaussian disorder correlation function (\ref{15}).
In summary, we have shown that the stable
self-energy matrix for $ {\cal D}(0) A > 1 $ has the continuous
replica symmetry broken form derived in Section VIC corresponding to the
VG phase. For
$ {\cal D}(0) A  < 1 $ we obtain that the full replica symmetric
solution derived in Section VIA is stable. This phase corresponds to the
vortex liquid VL.
The glass transition line between VG and VL is determined by
$ {\cal D}(0) A=1 $.

It is very difficult to solve the saddle point equation (\ref{210})
for a general disorder potential $ \hat{K}(q) $.
Nevertheless,   the glass transition line
$ {\cal D}_{\infty}(0) A_{\infty} =1 $ where
$ {\cal D}_{\infty} $ is defined in (\ref{34})
is valid in the general case.
We point out that equation (\ref{840}) leading
to the instability of the one-step replica symmetry-breaking
solution in the case
of the effective Gaussian disorder potential (\ref{15}) is 
still valid irrespective
of the correlation potential $ \hat{K}(q) $.
In general, the one-step
replica symmetry-breaking solution is stable if
\begin{equation}
 \eta_1[\tilde{\Delta}_1]\,  \kappa_1\left[\hat{K}\exp\left[-q^2
\, k_B T g(\Delta_1)/2 v
\right] \right] \le 1 \label{860}
\end{equation}
 with
\begin{align}
& \eta_1[\tilde{\Delta}]=  \tilde{\Delta}^{2}
\frac
{(1+5 \tilde{\Delta}/4)}
{(\tilde{\Delta}(1+\tilde{\Delta}/4))^{3/2}} \label{870} \\
&  \times
\left[\!\!4 \arcsin\left(\frac{\tilde{\Delta}^{1/2}}{2}\right)
-\left(\frac{\tilde{\Delta}}{1+ \tilde{\Delta}/4}\right)^{1/2}
\right]^{-1}        \,, \nonumber    \\
  & \kappa_1 \left[f\right] =
\frac{ \left[\int \frac{d^2q}{2\pi} \,q^4 f(q)\right]
\left[ \int \frac{d^2q}{2\pi}   f(q)\right] }
{2 \left[\int \frac{d^2q}{2\pi} \, q^2 f(q)\right]^2}
  \label{880}
\end{align}
where
we get an unstable one-step replica symmetry-breaking
 solution if and only if
$ \eta_1 \kappa_1 > 1 $.
Next, we consider the existence
of the continuous replica symmetry-breaking
solution.
For their existence  it was  crucial that in
(\ref{535}) the right hand side was larger than zero.
The corresponding equation without restrictions on
the correlation function is given by
\begin{equation}
 s \,({\cal D}_\infty(0) A_\infty)^{1/3}  =
\eta_2[\tilde{\Delta}(s)]
 \kappa_2 \left[\hat{K}\exp\left[-q^2  B[\Delta(s)]/2
 \right]\right]     \label{890}
\end{equation}
 with
\begin{align}
& \eta_2[\tilde{\Delta}(s)]=  \frac{\left[1+ \frac{5}{4}
\tilde{\Delta}(s)\right]^{5/3}}
{1+ \frac{2}{3}\tilde{\Delta}(s) + \frac{1}{6}\tilde{\Delta}^2(s)} \,,
\label{900}  \\
& \kappa_2 \left[f \right]
= \frac{2^{2/3}}{3}\frac{\left[\int \frac{d^2q}{2\pi}  \, q^6 f(q) \right]
f(0)^{1/3}}
{\left[\int \frac{d^2q}{2\pi}
 \, q^4 f(q) \right]^{4/3}}. \label{910}
\end{align}
To solve this equation for a general correlation function
$\hat {K} (q) $ without further approximations is not an easy task.
Nevertheless, the condition that a continuous replica
symmetry broken solution exist is given by the possibility
to solve (\ref{890}) for  $ s_c $ resulting in
\begin{equation}
 \eta_2[\tilde{\Delta}(s_c)]\,
\kappa_2 \! \left[\hat{K}\exp\left[-q^2  B[\Delta(s_c)]/2
 \right]\right] >  1  \,.   \label{915}
\end{equation}

Quantities like
$ \kappa_1, \kappa_2  $ are well known quantities
in probability theory.
The corresponding quantity called kurtosis measures
in one dimension the curvature of a probability function compared
to the Gaussian probability function. This is also valid for
(\ref{880}), (\ref{910}). The Gaussian correlation function (\ref{15})
has $ \kappa_1, \kappa_2= 1 $. A correlation function with a sharper tip and
longer tails like $ \sim \exp[-q^{\alpha}] $ for $ \alpha < 2 $
has  $ \kappa_1,\kappa_2 > 1 $. For $ \alpha > 2 $ which is more flat near
the origin with a shorter tail than the Gaussian function
has $\kappa_1,\kappa_2  < 1 $.

Next, we specialize the stability condition (\ref{860}) and (\ref{915})
to the vicinity of the glass transition line where
$ \tilde{\Delta}^{1/2} A_\infty /2     \ll 1 $.
By using (\ref{260}), (\ref{270}) and (\ref{570})
one can derive the validity of this condition by
$ (({\cal D}_\infty (0) A_\infty)^{1/3}-1)
\ll 1 $ irrespective of the disorder correlation function. In this regime,
we obtain by an expansion of $ \hat{K} $ around the origin which is
justified due to $ \tilde{\Delta}^{1/2} A_\infty/2    \ll 1 $
\begin{align}
& \kappa_1\left[\hat{K}\exp\left[-q^2 k_B T g(\Delta_1)/2 v \right]\right]
\approx  1 \!\! + \!
\frac{4 v^2}{(k_B T)^2 g^2(\Delta_1)} \label{920} \\
&
 \times \frac{1}{\hat{K}^2(0)}\left[\hat{K}(0)\left(
 \frac{\partial}{\partial (q^2)}\right)^2 \! \! \hat{K}(0)-
\left(\frac{\partial}{\partial (q^2)} \hat{K}(0)\right)^2\right] \,,
 \nonumber \\
& \kappa_2\left
[\hat{K}(0)\exp\left[-q^2 B(\Delta(s))\right]\right] \approx 1
+ \frac{8}{B^2(\Delta(s))}   \label{930} \\
&  \times
~\frac{1}{\hat{K}^2(0)}\left[\hat{K}(0)\left(
 \frac{\partial}{\partial (q^2)}\right)^2 \! \hat{K}(0)- \!
\left(\frac{\partial}{\partial (q^2)} \hat{K}(0)\right)^2 \! \right].
 \nonumber
\end{align}
One can derive the simple identity
\begin{equation}
 \kappa_1[K]  =  \hat{K}(0)\left(
 \frac{\partial}{\partial (q^2)}\right)^2 \hat{K}(0)\bigg/
\left(\frac{\partial}{\partial (q^2)} \hat{K}(0)\right)^2 \label{940}
\end{equation}
where $ \kappa_1[K] $ is the kurtosis (\ref{880})
built with the disorder correlation function $ K $  (\ref{11})
in position space. By using $ \eta_1[\tilde{\Delta}] \approx 1+20
\tilde{\Delta} /24 $ and  $ \eta_2[\tilde{\Delta}] \approx 1+17
\tilde{\Delta} /12 $
we obtain the simple  rules in Table I
in the regime $ (({\cal D}_{\infty}(0) A_\infty)^{1/3}-1)
\ll 1 $ for the stable saddle point
\begin{table}[h]
\begin{tabular} {c || c | c}
$ \kappa_1[K] $ & $ \le  1-20/6 A_\infty^2 $  & $   > 1-17/6 A_\infty^2 $ \\
\hline 
\rule[-1.3mm]{0cm}{0.5cm} 
saddle point & one-step breaking & continuous breaking \\
order of transition  &   third  order & third order  \\
\end{tabular}  \\
 \caption{Stable saddle points of (\ref{115}) as a function of 
the kurtosis $ \kappa_1(K) $ (\ref{880}) of the disorder correlation 
function $ K $ in real space. 
The second line of the table denotes the character of the stable 
solution of equation (\ref{210}). The third line stands for  
the order of the VG-VL transition. 
  \label{950}}
\end{table}
where $ A_\infty  $ has to be taken at the transition point.
We obtain from Table I a small transition region at
$ 1-20/6 A_\infty^2 \le  \kappa_1[K] \le  1-17/6 A_\infty^2 $ where
a higher-step replica symmetry broken  saddle point solution of (\ref{115})
should give the best free energy. We expect that this finite-step
replica symmetry-breaking solution leads also to  a third order 
glass transition in this range.

By using $ A_\infty = 4 c_{44} a^2 {\xi'_\infty}^2/ a_3 k_B T
\approx  b /2 \pi c_L^2 $ (see the discussion below (\ref{275}))
we obtain for the
high magnetic field part $ b \gg  0.5 $ of the glass transition line the
following simple result: \\
When the kurtosis $ \kappa_1 $ of the positional
disorder correlation function $ K $ is smaller
than the kurtosis of a
Gaussian function (flatter tip, shorter tail), the stable saddle
point solution of (\ref{210}) is given by a one-step replica symmetry
broken solution with free energy (\ref{280}) in the VG phase.
We obtain a third-order glass transition. When the kurtosis $ \kappa_1 $
is larger or equal  the kurtosis of a
Gaussian function then we have a continuous replica symmetry
broken solution with free energy (\ref{590}) in the VG phase in accordance 
with the one-step replica symmetric free energy (\ref{280}). 
According to Table \ref{950} we obtain that for lower magnetic fields
the border in the disorder function space of one-step replica symmetry 
 breaking solutions and continous replica symmetry 
breaking solutions moves to lower kurtosises.

\section{Solid phase}
In this section we determine the free energy
in the solid phase. This system corresponding to a string lattice
in a random potential was discussed in Ref.\,
\onlinecite{Giamarchi1}. Here, we reconsider it where we
took more emphasis on the determination of the free energy
of the vortex lattice in the low-temperature phase than the former work.
For $ F_{\rm var} $ we use again the approximation (\ref{140}).
For deriving this expression one has to consider other arguments
than in the fluid phase below (\ref{140}). First, we use
that the saddle point Green function calculated with (\ref{140})
fulfills $k_B T | {\bf G}_{\alpha \beta}(0)-
{\bf G}_{\alpha \beta}(a {\bf e}_i)|/v \ll a^2 $ justified in
Appendix~A. Here $ a {\bf e}_i $ is a nearest neighbor vector in the xy-plane.
Similar as in the considerations below (\ref{30})
we can  restrict us to the diagonal summands $ {\bf x}= {\bf x}' $
corresponding to (\ref{140}) where
non-diagonal terms being a factor
$ (B_{\alpha \beta }  + \xi^2)^{1/2}/ a $ smaller because as is shown in
Appendix~A $ B_{ \alpha \beta } \ll a^2 $ for almost the
whole range of replica indices.
  We point out that the
$ \alpha ,\beta  $-range where this inequality is not fulfilled
is important for
the long distance
behavior of the lattice fluctuations beyond the random manifold regime
\cite{Giamarchi1}
(see (\ref{1110}) below). Nevertheless, due to the  $ \alpha ,\beta
 $-sum in the
various free energy terms in (\ref{125}) one can show that these
contributions to the free energy are negligible where the
inequality  $ B_{\alpha \beta } \ll a^2 $ is not
fulfilled  (see the continuous version (\ref{170}) of (\ref{125}) and the
inequality (\ref{a2020})). Under the considerations above, we arrive at the
disorder Hamiltonian (\ref{140}) as the basic disorder Hamiltonian for
the solid low-temperature phase.

It is well known and can be also shown by a similar analysis as was done
for the fluid phase in the last section that finite-step replica symmetry
breaking solutions are unstable but the continuous replica broken
solution exist
which is stable as is shown by Carlucci {\it et al.} in Ref.\,
\onlinecite{Carlucci1} and Appendix~B.
This breaking of the replica symmetry corresponds to a glassy phase.
We now calculate this replica broken solution. The calculation is
similar to the calculation of the continuous
symmetry broken solution in the fluid
phase carried out in Section~VIC. As was done before for the fluid phase
we can restrict us to the transversal fluctuations in the displacement fields
 $ {\bf u}  $ because $ c_{11} \gg c_{66} $.
 Then we have to calculate $ g(\Delta) $
where we restrict us to the two lowest-order expansion terms
in $ \tilde{\Delta} \ll 1 $.
 By using (\ref{19}) we obtain
\begin{align}
& g(\Delta) \approx \frac{a_3^2}{c_{44}}  \frac{1}{ V_{\rm BZ}}
 \int_{\rm BZ}  d^2k dk_3
   \nonumber \\
&
 \frac{1}{[2-2 \cos(k_3 a_3)]+\frac{a_3^2 c_{66}}{a^2
 c_{44}} \sum_i [2-2 \cos(k_i a)]+ \tilde{\Delta}} \nonumber \\
&
\approx \frac{a_3^2}{c_{44}} \left(0.21-
\frac{\tilde{\Delta}^{1/2}}{16}\right) .   \label{1010}
\end{align}
Here we use the same approximation as in Ref.\,\onlinecite{Dietel2}
which means $ a_3^2 c_{66}/a^2 c_{44} \approx 4/ \pi $.

First, we determine the two solutions of (\ref{210}) corresponding to
(\ref{520}) and (\ref{535}). This results in
\begin{eqnarray}
 1. \quad \sigma'(s) &= & 0  \,,   \label{1020} \\
 2. \quad \tilde{\Delta}(s) & = &
 \frac{8}{3^{3/2}} ({\cal D}(0) A)^{1/2} s^{3/2} \,.\label{1030}
\end{eqnarray}
Instead of (\ref{550}) for $ s_c $ in the case of the fluid phase we find
for the solid phase
\begin{equation}
s_c^{1/2} = \frac{3^{1/2}}{2} \; {\cal D}^{2}(0) \; ({\cal D}(0) A)^{-3/2}
\,.
\label{1040}
\end{equation}
 This value was calculated by using (\ref{540}) with the approximation
 $ {\cal D}(2 B[\Delta(s_c)]) \approx {\cal D}(0) $ valid
 for $ c_L^2 \ll 1 $. This is correct in the vicinity of the melting line
 \cite{Dietel2}.
  Summarizing, we obtain for $ \tilde{\Delta}(s) $
  \begin{equation}
 \tilde{\Delta}(s)  = \left\{
 \begin{array}{ccc}
 \frac{8}{3^{3/2}} ({\cal D}(0) A)^{1/2} s^{3/2} & \mbox{for}
 &   s \le s_c \,, \\
  {\cal D}^6(0) ({\cal D}(0) A)^{-4}& \mbox{for} &
  s_c \le s \le 1   \,.
 \end{array}
 \right.     \label{1050}
 \end{equation}
From $ A \gg 1 $ near the melting line (see the remarks below
(\ref{275})) we obtain that $  \tilde{\Delta}(s)\ll 1 $ is in fact
fulfilled in the magnetic temperature regime, we are interested in
(note that $ {\cal D}(0) A \lesssim 1 $ on the melting
line for temperatures larger or in the vicinity of the
glass transition line).
Finally, we calculate the
free energy corresponding to (\ref{590})
by using (\ref{170}) with (\ref{1010}).
For $ \tilde{\Delta} \ll 1 $ we obtain
\begin{equation}
  \Delta f_{\rm var} \approx \frac{k_B T}{2} {\cal D}(0)
\bigg[1  -\frac{3}{20} {\cal D}^4(0)({\cal D}(0) A)^{-3} \bigg] \,.
\label{1060}
 \end{equation}
Here we neglect energy terms coming from the first term in (\ref{170})
corresponding to the kinetic part which is a factor $ \sim 1/10 $ smaller.
When comparing (\ref{1060}) with the free energy of the
quadratic disorder case (\ref{31}) of Section~IV  we obtain that only the
second term in (\ref{1060}) is different. This term should cancel the
first term in (\ref{1060}) for lower temperatures
resulting in a vanishing of the reentrant behavior of the BG-VG, BG-VL
line in the quadratic disorder case.

It can be seen from the derivation above and also Appendix~A that the actual form
of the self-energy function $ \tilde{\Delta}(s) $ depends on the
form of the disorder correlation function not only by one small
parameter. For $ s=s_c $ we have
$ B(\Delta[s_c]) \ll \xi'^2 $  but
for $ s \ll s_c $  we have $ B(\Delta[s_c]) \gg \xi'^2$ which means
that the form of the whole correlation potential is important when
solving the saddle point equation (\ref{210}).
This makes it difficult to solve this equation
in general. Nevertheless, for disorder correlation functions in the
vicinity of the
effective Gaussian disorder correlation function (\ref{15}) we think that
the result (\ref{1060}) should not be much changed.

\section{Observable Consequences}
Let us now apply the results obtained above
to find the BG-VG  line and the glass transition line of YBCO.
The entropy and magnetic induction jumps over the transition lines will
also be discussed.
We saw in Section~VII that the form of the
local  minimum of the variational free energy (\ref{115}) in
the high-temperature phase depends not on the
kurtosis of the disorder correlation function $ K $ where
the results are summarized in Table \ref{950}. Here the Gaussian disorder
correlation function with $ \kappa_1[K]=1 $
separates the regime
where we have a local minimum of $ F_{\rm var} $ of
the one-step symmetry-breaking form ($ \kappa_1[K] < 1 $) and of the
continuous  replica symmetry-breaking form 
($ \kappa_1[K] \ge 1 $) for large $ A_\infty $.

The free energies for both regimes coincide given by (\ref{280}) or 
(\ref{590}), respectively.
For the solid phase we obtain
the continuous symmetry-breaking solution (\ref{1060}) for the
Gaussian correlation function. For disorder correlation 
functions in the vicinity
of the Gaussian correlation function we can use this free energy as a first
approximation for the free energy of a general disorder
correlation function.
Taking into account (\ref{170}), (\ref{175}) we obtain
\begin{widetext}
 \begin{eqnarray}
 \Delta  f^{T \to 0}_{\rm var} & \approx &
  \frac{k_B T}{2} {\cal D}(0)
\bigg[ 1
  -\frac{3}{20} {\cal D}^4(0)({\cal D}(0) A)^{-3} \bigg]
\qquad  \qquad \qquad \qquad \qquad  {\rm BG \;  phase}   \,,\label{1070}   \\
 \Delta f^{T \to \infty}_{\rm var} & \approx &
 \frac{k_B T}{2} {\cal D}_{\infty} (0)  \left[1-
  ({\cal D}_\infty(0) A_\infty)^{-1/3}  \right]^3
\Theta[{\cal D}_\infty(0) A_\infty-1]
 \quad   {\rm VG-VL \;  phase}   \label{1080}
 \end{eqnarray}
\end{widetext}
in the regime near the melting and glass transition line.
The free energy is given by
$ F_{\rm fl} \approx  N f_{\rm var}=N(f_{\rm var}(0)+
\Delta f_{\rm var}) $ (\ref{170}), (\ref{175})
 where the disorder part of the free energy $ \Delta f_{\rm var} $
is given by (\ref{1070}) in the solid phase and
(\ref{1080}) in the fluid phase.
The intersection criterion corresponding
to (\ref{65}) in the quadratic approximation in
the disorder strength which determines the BG-VG, VG-VL line reads
\begin{align}
&   B_m \approx
 \frac{ \phi_0^5 \, (1-b)^{3} }{(k_BT)^2 \lambda^2_{ab}\lambda^2_c}
 \frac{ 3.9 \cdot 10^{-5} }{\pi^4} \, e^{-(2/k_B T)
(\Delta f^{T \to 0}_{\rm var }- \Delta f^{T \to \infty}_{\rm var})} 
  \nonumber \\
& ({\rm BG-VG, \;BG-VL\; line }) \,.  \label{1090}
\end{align}
Without disorder we have shown in Ref.\,\onlinecite{Dietel2} that
the melting criterion (\ref{85})
is equivalent to a Lindemann criterion where the Lindemann parameter
is given by $ c_L \approx 0.18 $. There are many papers which used
Lindemann-like criteria also to determine  the disorder induced
BG-VG  line
\cite{Ertas1,Mikitik1,Menon1,Radzyner2,Giamarchi2}. In Ref.\,
\onlinecite{Mikitik2} Mikitik and Brandt even tried to derive a Lindemann-like
criterion for the BG-VG, VG-VL line from an intersection
criterion similar to the one
used here. Because these Lindemann-like rules do not look similar to our
microscopically derived melting criterion (\ref{1090})
we do not try to go further in this direction.

\begin{figure}[t]
\begin{center}
\includegraphics[height=9cm,width=8.5cm]{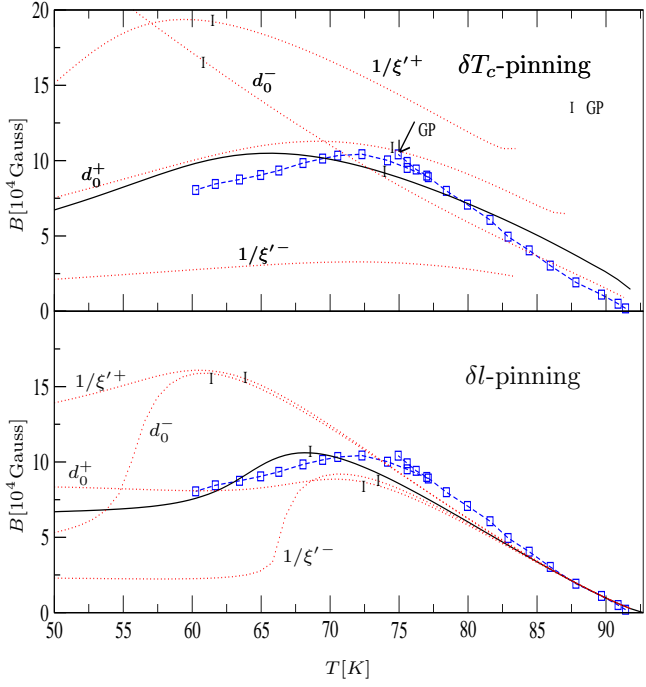}
\end{center}
 \caption{The BG-VG, BG-VL first-order transition lines
$B_m(T) $ given in (\ref{1090}) for $ \delta T_c $-pinning (upper figure)
and  $ \delta l $-pinning  (lower figure). The solid lines are calculated
with parameters for $ d_0 $ and $ \xi' $ which gives one of the best
fits to the experimentally determined \cite{Bouquet1} BG-VL line
(square points) within the
pinning mechanism
($ 2\pi d_0 \, \xi_{ab}^2/{\xi'}^2=1.5 \cdot 10^{-7} $ and $ \xi_{ab}/\xi'= 1.59 $ for
$ \delta T_c$-pinning,
$ 2\pi d_0 \, \xi_{ab}^2/{\xi'}^2=1.32 \cdot 10^{-6} $ and $ \xi_{ab}/\xi'= 1.49 $
for $ \delta l $-pinning).
Dotted curves are calculated by a variation of
these parameters given by
$ d^\pm_0=(1 \pm 1/2) d_0 $ and  $ 1/{\xi'}^\pm=
(1 \pm 1/2)^{1/2}/\xi'  $ where only
one parameter was varied.
We wrote that parameter
at the curve.
The vertical markers denote the
intersection points of the glass transition line and the
BG-VG, BG-VL line named GP. From Clausius-Clapeyron equation (\ref{2010})
the critical points CPs
are  determined by an extremum of the BG-VG, BG-VL lines}
\vspace*{0cm}
\end{figure}

\begin{figure}[t]
\begin{center}
\includegraphics[height=6cm,width=8.5cm]{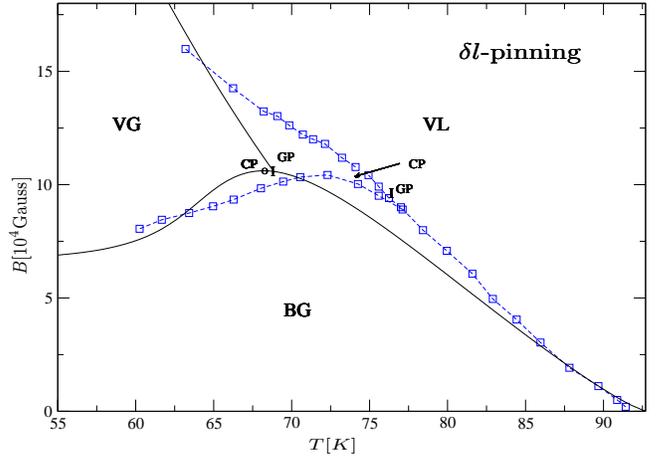}
\end{center}
 \caption{Phase diagram for YBCO. Solid lines represent the
 theoretical determined phase transition lines between the various phases
calculated for $ \delta l $-pinning with
 $ 2\pi d_0 \, \xi_{ab}^2/{\xi'}^2=1.32 \cdot 10^{-6} $ and
$ \xi_{ab}/\xi'= 1.49 $ corresponding to 
the solid line in the lower picture in Fig.~4.
 The glass transition line VG-VL was calculated
  from (\ref{1100}). Square points represent the
 experimentally determined phase diagram of Bouquet  {\it et al.}
 \cite{Bouquet1}.}
\vspace*{0cm}
\end{figure}

As derived in section VIB,
the glass transition line which is the border of the
replica symmetric solution of the M\'ezard-Parisi variational calculation
and the one-step  replica symmetry-breaking solution, where the stabilities was
discussed in section VII,  is determined by
$ m_1 = 1 $ in (\ref{260})  resulting in
\begin{equation}
{\cal D}_\infty (0) A_\infty = 1    \qquad     ({\rm VG -VL  \; line})  .     \label{1100}
\end{equation}
This equation corresponds to the depinning temperature $ T_{\rm dp} $
of  a one-dimensional string in three dimensions
in a random environment \cite{Blatter1}. The Larkin length
 $ L_{c} $ is defined by the length where we have a coherently pinning
of the string which means $ u^2(0,L_c)  =\xi'^2 \sim \xi^2_{ab} $ with
\begin{equation}
u^2(L,L_3)\equiv  \overline{\langle ({\rm u}(L,L_3)-
{\rm u}(0,0))^2 \rangle}   \,.  \label{1110}
\end{equation}
When temperature fluctuations become larger there is
a softening of the impurity potential which is important for the
length of the coherently pinned vortices. This correction is
important when this fluctuation length becomes equal to $ L_c $
calculated for $ T=0 $. This depinning
temperature is given by (\ref{1100}) \cite{Blatter1}.
We mention that it is difficult to distinguish
experimentally by diffraction
 experiments in which of the two
classes $ \kappa_1[K] > 1-17/6 A_\infty^2 $ or  $ \kappa_1[K]
\le 1-20/6 A_\infty^2 $
the  disorder correlation potential $ K $ of a given experiment
belongs.  In both regimes we obtain $ u^2(0,L_3) \propto
(k_B T) L_3 /c_{44} a^2    $
in the VG-VL phase (the proportionality constant is different for
both regimes, see also the discussion in Appendix~A).
This is reasoned in  the vanishing support of $ \sigma(s) $ in the
vicinity of the origin, fact for the
the one-step replica symmetry-breaking regime $ \kappa_1[K] \le
1-20/6
A_\infty^2 $ (\ref{235})
as well as the continuous replica symmetry-breaking regime
$ \kappa_1[K] >  1-17/6 A_\infty^2 $ (\ref{530}) and (\ref{570}). It means
that thermal fluctuations are dominant over disorder fluctuations.
Note that $  u^2(L,L_3) $ for $ L \not =0 $ diverges in the
VG as well as the VL phase characteristic for defect dominated
phases also seen
before for the system without disorder.
For the BG-phase in the random manifold regime
we obtain $  u^2(L,L_3) \propto (L^2+L^2_3)^{1/6} $
in accordance with former calculations \cite{Giamarchi1}.
The derivation beyond the random manifold regime where the lattice
structure is important leads to $  u^2(L,L_3) \propto \log(L^2+L^2_3) $
\cite{Giamarchi1}.

Mikitik and Brandt found in Ref.\,\onlinecite{Mikitik1, Mikitik2}
that their analytical derived BG-VG, BG-VL curve is a function of
the Ginzburg number $ {\rm Gi} =32 \pi^4
(\lambda_{ab}(0) \lambda_c(0) T_c/\phi_0^2 \xi_{ab}(0))^2 $, $ b$, $T/T_c$,
and the disorder function  $ {\cal D}(0) $. This can be
also shown easily for the melting curve (\ref{1090}). Note that the
disorder constant $ D $ in
Ref.\,\onlinecite{Mikitik1,Mikitik2} is a function of $ b$, $T/T_c$,
$ {\rm Gi} $ and our disorder function $ {\cal D}(0) $.

In Fig.~4 we show the BG-VG, BG-VL curves given by (\ref{1090})
for various values $d_0$, $ \xi' $. 
The upper curves are calculated
with a $ \delta T_c $-pinning correlation function (\ref{12}), the lower curves
for a $ \delta l $-pinning impurity correlation function (\ref{13}).
For clearance we do not show the critical points CP
on the melting line in the figure which are characterized
by zero entropy jumps $ \Delta S_l $ per double layer and vortex
over the transition line. These points can be easily
marked in the figure since they
correspond to the extrema of the melting line $ B_m $ due to the
Clausius-Clapeyron equation given in (\ref{2010}) below.

The intersection point of the glass transition line BG-VL which is
calculated by (\ref{1100}) with the
BG-VG, BG-VL line is denoted by GP in the figure. The square
points with the dashed line denotes the experimentally determined
BG-VG, BG-VL line of Bouquet {\it et al.} \cite{Bouquet1}
shown also in Fig.~2 for comparison.
In the $ \delta T_c $ part of the figure, we find no solutions
for the equation (\ref{1090}) near $ T \approx T_c $. The parameters
of the straight lines
in the figure are chosen in such a way that we reproduce in one of
the best ways the form of the experimentally melting line of
Bouquet {\it et al.} \cite{Bouquet1} and also the position
of the experimentally found
CP and GP. These experimentally chosen parameters are
$ 2\pi d_0 \, \xi_{ab}^2/{\xi'}^2=1.32 \cdot 10^{-6} $ and $ \xi_{ab}/\xi'= 1.49 $
for $ \delta l $-pinning,
$ 2\pi d_0 \, \xi_{ab}^2/{\xi'}^2=1.5 \cdot 10^{-7} $ and $ \xi_{ab}/\xi'= 1.59 $ for
$ \delta T_c $-pinning. Thus, we obtain
that the correlation length $ \xi' $ of the
disorder potential almost corresponds to the coherence length
$ \xi_{ab} $ of the superconductor.
The reason that  $ \xi_{ab}/\xi' $ is larger than one could be due to lattice
influences on the effective broadening of the vortex (see the notes below
(\ref{13})).
Finally, we mention the similarity of the $ d_0 $ parameter
values in the Parisi case
and the corresponding values in the quadratic disorder case of
Section~IV.

The curves of representative variations of these almost
optimal parameter values are shown by the
dotted curves. We obtain from the figure
as was also the case in the second-order perturbative discussion in
Section~IV that the $ \delta T_c $-pinning curves fits less to the
experiment than the $ \delta l $-pinning curves. This comes mainly from the
smoothness of the disorder parameter $ d(T) $ in (\ref{13})
as a function of $ T $ resulting in the slow variation of the transition line
$ B_m(T) $ seen in the upper part of Fig.~4.
From Fig.~4 we obtain that the glass
intersection point GP and the critical point CP does in generally not
coincide. This was just mentioned in Ref.\,\onlinecite{Beidenkopf1} for
BSCCO where in the experiments this difference is not seen yet
maybe because of experimental uncertainties.

One of the most interesting results of our calculation is that
the reentrant behavior of the melting line and the experimentally not
seen low $ B $ parts of the BG-VG curves in the quadratic disorder
calculation of section IV (see Figs.~2 and 3)  vanished in the Parisi
approach. It is remarkably that the
large descend of the curves in the direction to lower  temperatures
in Fig. 2 within the quadratic approach
are  smoothed within the M\'ezard-Parisi
approach such that the BG-VG transition curves is almost
horizontal. There are various
forms of the BG-VG lines in the literature.
One of the reasons of the differences in  the
various experiments comes presumably from the strong
dependence of the BG-VG line on the depinning function
$ d(T) $ via an exponential behavior
in (\ref{1090}). Any perturbational effects like surface effects or twinning
areas in the crystal can change the functional form of the curve
at small temperatures easily. We note that especially the
strong dependence of the BG-VG curve on small variations of the disorder
correlation length $ \xi' $ having its reason in the quadratic dependence of
$ \xi' $ in $ A $ (\ref{275}) which is contained as a third-order
summand in the free energy of the solid phase in (\ref{1060}). The form
of the free energy in the solid phase is the most dominant factor for the
form of the BG-VG curve in the vicinity of the critical point CP.
This is in contrast to
the free energy in the high-temperature phase which just
grows in importance  beyond the glass intersection point GP but is still
small compared to the free energy part of the solid phase.

In Fig.~5 we show the whole phase diagram for
the parameters of the solid curve in the lower $ \delta l $-pinning part
of Fig.~4 (solid curve).
For comparison (square points with dashed curve) we show
also the experimentally determined phase diagram of Bouquet {\it et al.}
of Ref.\,\onlinecite{Bouquet1}.
Both phase diagrams look rather similar except that the CP and GP
points of the theoretical determined phase diagram lies a little bit
lower in temperature in comparison to the experimental ones.
The upper curve between the VG-VL phases show the glass transition
line calculated by (\ref{1100}). There is a small discrepancy in the slope
of the line between theory and experiment.
Note that for $ \tilde{\lambda} \approx \lambda $ which is the
case for BSCCO \cite{Brandt1} we get that
$ {\cal D}_{\infty}(0) A_{\infty} $ is in fact independent
of the magnetic field $ B $
resulting in a vertical glass transition line. This is in good accordance
to the experiments \cite{Beidenkopf1}.

Next, we calculate the entropy and magnetic induction jumps over the
BG-VG, BG-VL first-order line.
Denoting the spacing between the CuO$_2$ double layers
 by $ a_s $  we obtain
for the entropy jump per double layer and vortex over the BG-VG, BG-VL line 
\begin{equation}
\Delta S_l  \approx   k_B T_m \frac{a_s}{a_3} \frac{\partial}{\partial T_m}
\, \ln[Z^{T \to \infty}_{\rm fl}/Z^{T \to 0}_{\rm fl}]     \label{2000}
\end{equation}
and a corresponding equation for the glass transition line.
Now we make use of the Clausius-Clapeyron equation
which relates the jump of the entropy density of
a first-order transition line
 to the jump
of the magnetic induction
by
\begin{equation}
 \frac{a_3 \Delta S_l}{v a_s} =-\frac{d H_m}{dT}
\frac{\Delta B}{4 \pi} \label{2010}.
\end{equation}
Here  $ H_m $ is the
external magnetic field on the BG-VG, BG-VL line. Because
$ B \sim H_{c2}(T) $ for YBCO we can use  $ H \approx B $ in the
Clausius-Clapeyron equation (\ref{2010}). Equation (\ref{2010})
is not appropriate
for a numerical evaluation because of the vanishing of the denominator at
the saddle points of the BG-VG, BG-VL line which are canceled due
to zero points in the numerator. By using the intersection criterion for
the transition line we can transform (\ref{2010}) to
\begin{equation}
\Delta B  \approx   k_B T_m \frac{4 \pi }{v}
\frac{\partial}{\partial B_m}
\, \ln[Z^{T \to \infty}_{\rm fl}/Z^{T \to 0}_{\rm fl}] \,.    \label{2015}
\end{equation}
This equation can be also derived from thermodynamical relations
under the  considerations $ \Delta B / B_m \ll 1 $ which
we also used by taking the intersection criterion for the free energy
and not for the corresponding Gibb's potential in this paper
\cite{Tinkham1}.

\begin{figure}[t]
\begin{center}
\includegraphics[height=8.0cm,width=8.5cm]{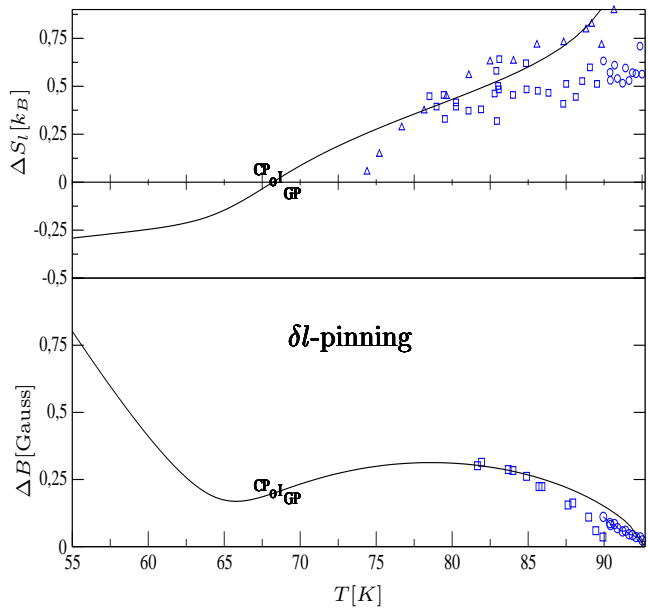}
\end{center}
 \caption{In the upper figure
we show the entropy jump $ \Delta S_l $  per double layer and vortex
according to (\ref{2000}). The points in the figure are
entropy jumps determined by experiments
(circles \cite{Willemin1}, squares \cite{Schilling1},
triangles \cite{Bouquet1}). In the lower figure  we show
the magnetic induction jumps
$ \Delta B $ calculated by the help of (\ref{2015}).
Experimental points in this figure  are from Ref.\,
\onlinecite{Willemin1} (circles) and Ref.~\onlinecite{Welp1}
(squares).
The solid curves in both figures
correspond to the jumps over
the BG-VG, BG-VL line.
We used for the whole figure parameter values
 $ 2\pi d_0 \, \xi_{ab}^2/{\xi'}^2=
1.32 \cdot 10^{-6} $ and $ \xi_{ab}/\xi'= 1.49 $
in correspondence to the parameter values in Fig.~5.
} \vspace*{0cm}
\end{figure}

In Fig.~6 we show
$  \Delta S_l $ and $ \Delta B $ for the parameters used in Fig.~5
over both lines.
We show in the upper part of the figure
$ \Delta S_l $ with experimental points
of various torque and SQUID experiments
(circles \cite{Willemin1}, squares \cite{Schilling1},
triangles \cite{Bouquet1}) for the entropy jump over the BG-VG,
BG-VL line.
For the parameters used in Fig.~5  we obtain a value for the CP of $68 K $.
In the lower part of Fig.~6  we show the magnetic induction jumps
$ \Delta B $.
The square and circle points are experiments
(circles \cite{Willemin1}, squares \cite{Welp1}).
Finally, we note that our curves of the entropy and magnetic
induction jumps over the BG-VG, BG-VL line
is in qualitative agreement with similar curves calculated
within the Ginzburg-Landau approach for YBCO by Li and Rosenstein \cite{Li1}.
The main difference is that they obtain a zero point in the magnetic induction
jump curve in the vicinity of the critical point which has its reason in
the reentrant behavior of their calculated  melting line to second order
in the disorder potential (see (\ref{2010}) by taking
into account that $ d B_m/ dT $ is infinite at the reentrant
points). We expect, as was also the
case in the elasticity approach used here, that this zero point
 vanishes when going
beyond second-order perturbation theory leading to the vanishing of the
reentrant behavior. \\[0.5cm]

Finally, we come back to a discussion of the scenarios of the phase diagram
for YBCO given in the introduction of this paper.
We did not find a slush phase within our numerical
examinations of equation (\ref{1090}) during this work irrespective
of the parameter range. This in accordance to
the Ginzburg-Landau calculations of Li and Rosenstein in Ref.\,
\onlinecite{Li1}. This means that our phase diagram is only in accordance
with the second scenario of a unified BG-VG, BG-VL first-order line discussed
in the introduction of this paper. Due to the controversy of
this  phase we cannot determine within our theoretical approach
whether it  is in fact existent or not. It was claimed in the experimental
paper \cite{Nishizaki1} that the slush phase only exists within a really
small doping region where the entropy jumps over the first-order line
between the slush phase VS and the vortex liquid VL is two orders smaller
than the entropy jumps over  the BG-VL melting line. The intersection
criterion of the high and low-temperature free energy used in this paper
by a perturbative calculation in both phases uses
the assumption that the slope difference corresponding to the entropy jumps
is not too small. This could be the reason that we do not see the slush
phase. We point out
that to our knowledge there exist no theoretical model
which shows without doubt the existence of this phase.

One of the main findings in this work for  
vortex model (\ref{2}) is that the order of the glass 
transition VG-VL is of third order irrespective of the form of the 
disorder correlation potential. 
We point out that a third order 
phase transition having a smooth 
heat capacity should show scaling behavior with a non-trivial fix 
point in a renormalization group calculation.  
Prominent examples 
of third order phase transitions  
is the non-interacting homogeneous three dimensional 
Bose gas across the Bose Einstein transition \cite{Junod1} or the 
large $ N' $-limit of the two dimensional $ U(N') $ lattice gauge theory 
with a variation in the coupling constant \cite{Gross1}.   
It was noticed  
in Ref.~\cite{Junod1} that the heat capacity curves for BSCCO over 
the superconducting transition without magnetic field looks 
rather similar to the heat capacity curves of the homogeneous Bose gas.   
For YBCO this transition looks more like the $ \lambda $-transition 
of $^4$He. A discussion of scaling relations in higher order phase 
transitions and their classification due to Ehrenfest 
can be found in Ref.~\onlinecite{Kumar1}. 
  
Fisher {\it et al.} proposed in Ref.~\onlinecite{Fisher1} a 
scaling behavior of the 
VG-VL glass transition  where they introduce a disorder phase correlation 
length $ \xi_G $ with scaling 
$ \xi_G \sim  |T-T_G|^{-\nu} $ in the fluid phase near  
glass 
transition temperature $ T_G $ on the VG-VL transition line. 
This scaling proposal  
was later on approved experimentally via measurements of 
the current voltage characteristics over the transition region \cite{Gammel1}. 
There are now a number of experiments \cite{Gammel1,Klein1} 
and computer simulations \cite{Reger1,Lidmar1,Kawamura1} of 
various models for  superconductors showing also this 
 scaling behavior where in most cases the disorder phase 
correlation exponent lies in between $ 0.8 \le \nu  \le 1.7$. 
One can connect the phase correlation scaling exponent $ \nu $ 
with the heat capacity exponent $ \alpha $ defined by 
$ C\sim  |T-T_G|^{-\alpha} $ where $ C $ is the heat capacity via the 
hyperscaling relation $ \nu d=2 -\alpha $. Here $ d $ is the 
dimension of the system which means $ d=3 $ in our case. Thus, most 
of the experimentally determined and computer simulated systems have 
an $ \alpha $-exponent lying in between $ -3.1 \le \alpha \le -0.4 $. 
This corresponds to  a phase transition of order three or even higher 
within the Ehrenfest definition of phase transitions \cite{Kumar1}.

\section{Summary}
In this paper, we have derived the phase diagram for
superconductors having their phase transition lines at
high magnetic fields near $ H_{c2} $, such as YBCO.
The aim was to obtain a unified analytic theory for the BG-VG, BG-VL
transition as well as for the glass transition lines.
The model
 consists of the elastic degrees  of freedom
of the vortices with additional defect fields describing in the most
simple way the defect degrees of freedom of the vortex lattice.
For the  impurity potential we restricted
us to weak pinning  $\delta T_c $ and $ \delta l $-correlated impurities
\cite{Blatter1}.

First, we have derived the effective low- (\ref{19}) and high-temperature
Hamiltonians (\ref{25}) without disorder in Section~III.
The low-temperature Hamiltonian
consist of the well-known elastic Hamiltonian of a vortex lattice
where defects are frozen out.
At high-temperatures, the stress fields
are frozen out leading to the high-temperature Hamiltonian (\ref{25}).
In Section~IV we have
carried  out
the disorder averaging to
second-order perturbation theory
with these low- and high-temperature Hamiltonians to find the
BG-VG, BG-VL transition line by the application of the
intersection criterion.
The result given in Eq.~(\ref{65})
and displayed in
 Fig.~2
shows a  reentrant behavior.
The low-${B}$
 behavior
of the calculated
transition line was not in agreement
with  experiment. This led us
to  calculate the free energy in the low- and
high-temperature phases
using the non-perturbative approach
of M\'ezard-Parisi. In Section~VI we calculated
the variational free energy
in the high-temperature liquid phase.
We obtain a glass transition
from a replica symmetric solution corresponding to the vortex liquid
VL to a symmetry broken solution corresponding to the vortex glass phase
VG. The position of the glass transition line
fulfills equation (\ref{1100}) describing
 the depinning  transition of a
string with stiffness  $ c_{44} a^2 $ in three dimensions.
The degree of
replica symmetry breaking of the variational Hamiltonian
depends on the form of the
disorder correlation function.
The high-temperature part of the free energy is given by (\ref{1080}).
 For high magnetic fields near $ H_{c2} $
we got the following result:
We obtain a one-step  replica symmetry-breaking solution
when the kurtosis $ \kappa_1 $
of the disorder correlation function in position
space defined in (\ref{880}) is smaller than one. 
In the case that the kurtosis is larger than or equal to one
we obtain a full replica symmetry broken solution.
In both cases we obtain a third order glass 
transition line. The Gaussian correlation function is the border in 
the disorder correlation function space with $ \kappa_1=1 $.  
Corrections to this simple rule relevant for lower magnetic fields
are given in Table \ref{950}.

In Section~VIII,
we calculate the free energy of the vortex system in  the low-temperature
solid phase (BG), given by (\ref{1070}).
 The stationary solution for the self-energy
matrix in replica space is continuous replica symmetry broken.
By using the intersection criterion for the low- and high-temperature
free energies, we calculate the expression for the  unified  BG-VG, BG-VL line
given by (\ref{1090}). In Fig.~4 we show the unified BG-VG, BG-VL line
for various parameters for both pinning mechanisms.
We obtain that $ \delta l $-pinning  fits  much better to the
experiments than $ \delta T_c $-pinning. It is seen that the
reentrant behavior of the second-order perturbation theory carried out in
Section III  vanished in this non-perturbative approach.
In Fig.~5, we show the theoretical determined phase diagram for YBCO.
Fig.~6 shows  the entropy jumps and magnetic field jumps over the
BG-VG, BG-VL transition line.
Finally, we calculated 
heat capacity scaling exponents $ \alpha $ 
from disorder phase correlation exponents $ \nu $ 
determined from  experiments and computer
simulations via the hypercaling relation across the glass transition line 
VG-VL which is only consistent with 
a third or even higher order VG-VL phase transition line.


\begin{appendix}
\section{Justifications for approximation of
disorder Hamiltonian (\ref{140})}
We restrict us here to the case of transversal fluctuation where the
generalization to arbritrary fluctuations is straight forward.
That (\ref{140}) is valid for the high-temperature fluid
phase was shown below (\ref{163}).

In the solid phase, we first have to show that
$ k_B T \, 2\,  |G^T_{s}(a{e}_i)-G^T_{s}(0)|/v  \ll a^2 $
in the interesting regime near the melting line where
$ G^T_{s} $ is the full Green function.
$ a {\bf e}_i $ is a nearest neighbor vector in the xy-plane and
$ s $ is a continuous Parisi index.
It follows from \cite{Mezard1}
\begin{align}
& G^T_s({\bf x})=  \frac{1}{V_{\rm BZ}} \int_{\rm BZ} d^2k dk_3
\; e^{i {\bf k} \cdot {\bf x}} \; G_0^T
\label{a2010}\\
& \times  \left[
\frac{1}{s} \frac{\Delta(s)}{({G_0^T})^{-1}+\Delta(s)}
+\int_0^s \frac{ds}{s^2}
\frac{\Delta(s)}{({G_0^T})^{-1}+\Delta(s)}\right] \nonumber
\end{align}
that the nearest neighbor fluctuations are in fact much
smaller than the nearest neighbor distance $ a $
when taking into account (\ref{1010}), (\ref{1050}),
$ c^2_L \ll 1 $ and   $ D(A)  A \lesssim  1 $ near the melting and glass
line which is the regime we are interested in.

Finally, we have to show that $ B(\Delta(s)) \ll a^2  $ for almost all
$ s > 0$. From (\ref{200}),  (\ref{1050}) with (\ref{1010}),
$ c_L^2 \ll 1 $ and $ D(0) A \lesssim 1 $  we obtain  $ B(\Delta(s)) \ll a^2$
for
\begin{equation}
s \gg   c_L^8 \,.
  \label{a2020}
\end{equation}
which is almost the whole $ s $-region. The extreme small
range $ s \ll   c_L^8  $
has no relevance for the free energy result.
As mentioned above, this small  $ s $-region of $ \sigma(s) $
 becomes relevant only when  calculating disorder fluctuations
(\ref{1110}) beyond the random manifold regime which corresponds to
distances $ L $, $ L_3 $ where $ u^2(L,L_3) \gg a^2 $
\cite{Giamarchi1}.

\section{Stability of M\'ezard-Parisi
solutions}
In this section, we consider the stability criterion of the
M\'ezard-Parisi theory in the large $ N' $-limit and in
the Bogoliubov variational method.
First, we  reconsider the derivations of Carlucci {\it et al.}
\cite{Carlucci1} for
the stability conditions in the case of the large
$N'$-limit. Then we derive the corresponding stability
criteria in the variational
approach considered in Section~VI. To our knowledge this was not done
before in the literature.

In order to compare the vortex lattice theory with two component
displacement fields with the $ N'=2 $ isotropic random manifold
theory of M\'ezard and Parisi we restrict us in the following first
to the transversal displacement fields justified above as a good approximation
in both phases. A generalization to the full fluctuations
is straight forward.
 The difference
of the stationary and stability expressions
in both phases for the vortex lattice
and the isotropic $ N'=2 $ random manifold theory of M\'ezard and Parisi
comes then mainly  due to a difference in the kinetic part of the
Hamiltonian $ G_{0}^{-1} $ in Eq. (\ref{130}).
As was shown in Ref.\,\onlinecite{Mezard1} the saddle point equation
of both approaches looks rather similar except that in the
large $ N' $-limit the saddle point
equation (\ref{210}), $ {\cal D}_0(x)  $ should be replaced by
 $ f(x) $ where $ f(x)=\Delta(\sqrt{x})/ (K_B T)^2  $
(we take the reversed sign to the M\'ezard-Parisi
definition).
$ \Delta(\sqrt{x}) $ is the impurity correlation
function in (\ref{90}).
To derive this, M\'ezard and Parisi
insert in the action of
the isotropic random manifold system, auxiliary fields. By integrating out
the fluctuating displacement fields of the random manifold the large
$ N'$-limit corresponds to a saddle point approximation in the
auxiliary fields. This results in the saddle point equation (\ref{90}).
By the definition of
\begin{equation}
\tilde{f}(x) =\int_0^\infty d\alpha \;e^{-\alpha} f(\alpha x)
\label{a10}
\end{equation}
M\'ezard and Parisi obtain (\ref{210}) for the general variational
approach where the disorder function
$ {\cal D}_0(x)  $ is replaced by
$ \tilde{f}(x) $.

\subsection{Stability
in the large $ N' $-limit approach  of M\'ezard
and Parisi}
The stability of the stationary solution
(\ref{210}) comes from the stability of the
saddle point approximation of the action in the auxiliary fields.
This results in the stability matrix \cite{Mezard1} (we take into account
only the less stable part of the stability matrix corresponding to zero
moments)
 \begin{equation}
M^{\alpha \beta, \gamma \delta}
 = \frac{1}{2 f''(L^{(1)}_{\alpha \beta, \alpha \beta}) }
\delta_{\alpha \beta,\gamma \delta}
 - L^{(2)}_{\alpha \beta,  \gamma \delta} \label{a20}
 \end{equation}
with
\begin{align}
L^{(1)}_{\alpha \beta,\gamma \delta}
&  =  \frac{(k_B T)}{v \, V_{\rm BZ}}
 \int_{\rm BZ} d^2k dk_3
 \; (G_{\alpha \gamma }-G_{\alpha \delta}-G_{\beta \gamma}+
G_{\beta \delta})               \label{a22}  \\
L^{(2)}_{\alpha \beta,\gamma \delta}
& =   \frac{(k_B T)^2}{v^2 \, V_{\rm BZ}}
 \int_{\rm BZ} d^2k dk_3
 \; (G_{\alpha \gamma }-G_{\alpha \delta}-G_{\beta \gamma}+
G_{\beta \delta})^2               \label{a25}
\end{align}
and  $ \alpha < \beta $ and $ \gamma < \delta $.
Here $ G_{\alpha \beta} ({\bf k}) $ stands for the transversal component
of the Green function (\ref{130}) in the case of the vortex lattice
or the corresponding Green function in the case of the isotropic
two component $ N'=2 $ random manifold system \cite{Mezard1}.
The stability of the saddle point of (\ref{125}) fulfilling the
discrete version of the self-energy equation (\ref{210}) is given when
all eigenvalues of the stability matrix (\ref{a20}) are positive.

$ M^{\alpha \beta, \gamma \delta} $ is a four index ultrametric matrix
\cite{Dotsenko1}.
It was
shown by Kondor {\it et al.} \cite{Kondor1} and later on by
Temesv\'{a}ri {\it et al.}
\cite{Temesvari1} that one can divide the eigenvalues of
matrices of the form (\ref{a20}) in three classes.
The first two families consist of vectors in  the {\it longitudinal sector}
of dimension $ R+1 $ and $ R $
{\it anomalous sectors} of dimension $ R+1  $ depending
explicitly on the form of
the ultrametric matrix.
Here, we denote $ R $ by the level of hierarchy of the self-energy
matrix $ \sigma_{\alpha \beta} $ fulfilling the
stationarity condition (\ref{210}). This means
$ R=0 $ for the replica symmetric solution
calculated in section VIA  and $ R=1 $ for the one-step solution
given in section VIB for the fluid phase.

There is no closed form in the literature
for the eigenvectors and eigenvalues of the matrix in (\ref{a20}) for the
first two families.
Nevertheless, it is able to block diagonalize the matrix
$ \underline{\rm M}$
given by (\ref{a20}) in the
various sectors
\cite{Temesvari1}.
Following Temesv\'{a}ri {\it et al.}
\cite{Temesvari1}, we denote the size of the
Parisi blocks as $ p_r $, $ r=1\ldots R$, where $ R $ is the maximum level
of replica symmetry breaking. We denote $ p_0=n $ and
$ p_{R+1}=1$, the latter being the size
of diagonal elements. The matrix elements $ \sigma_{\alpha \beta} $, that
belong to the $r$th level of replica symmetry breaking
are all equal to a number denoted by
$ \sigma_r $, $ r=0,\ldots,R $. The replica overlap function  is defined by
$ \alpha \cap \beta =r $ when $ \sigma_{\alpha \beta}= \sigma_r $.

Denoting $ u^k_r $ with $ 0 \le r \le R$
the basis vectors in the first two families.
For $ k=0 $ which is the longitudinal sector we obtain for the
$ R+1 $ basis vectors \cite{Temesvari1}
\begin{equation}
(u^0_r)_{\alpha \beta}= \left\{
\begin{array}{ccc}
1 & \mbox{for} &  \alpha \cap \beta=r \,,  \\
0 & \mbox{for} &  \alpha \cap \beta \not=r \,.
\end{array}   \right.   \label{a30}
\end{equation}
The basis vectors  $ u^k_r $
for $ k \not=0 $ corresponding to the anomalous sector
 can be found in Ref.\,\onlinecite{Temesvari1}.

The third family of eigenvectors of ultrametric matrices as for example
(\ref{a20})  is named the  {\it replicon sector}. It consist on several
one-dimensional subfamilies labeled by $ r=0, \ldots, R $
and $ k,l= r+1, \ldots, R+1 $.
The corresponding one-dimensional subspaces are eigenspaces with the
eigenvalues denoted by $ \lambda(r;k,l) $. The eigenvectors
corresponding to the basis vectors in this sector can be found in
Ref.\,\onlinecite{Temesvari1}. Note  that these eigenvectors
do not depend on the entries of the ultrametric matrix.
The eigenvalues $ \lambda(r;k,l) $
can be generally expressed via the matrix elements of the ultrametric
matrix \cite{Temesvari1}.
In the case of the concrete
ultrametric matrix $ \underline{\rm M} $ (\ref{a20})
one finds \cite{Carlucci1}
\begin{equation}
\lambda(r; k,l)=
\frac{1}{2 f''(2 (K_B T /v) (g_{R+1}-g_r))}
 - L'_{kl}  \label{a40}
 \end{equation}
with
\begin{align}
& L'_{kl}=2 \frac{ (k_B T)^2 }{v^2\, V_{\rm BZ}}  \label{a45}  \\
& \times \int_{\rm BZ} d^2k dk_3
 \; \frac{1}{[(G_0)^{-1}+\Delta_{l-1}]}  \;
\frac{1}{[(G_0)^{-1}+\Delta_{k-1}]} \,.  \nonumber
\end{align}
Here $ g_{k} $ corresponds to the value of the transversal
component of the Green function $ G_{\alpha \beta} $ (\ref{130})
integrated over the momenta as in (\ref{180}) or the
corresponding random manifold Green function
with $ \alpha \cap \beta=k $. The eigenvalues $ \Lambda(r;k,l) $ for
$ k=l=r+1 $ are the most singular ones for definite $ r $.
One can show easily \cite{Carlucci1} that these most singular
eigenvalues are zero in the case of continuous symmetry-breaking
solutions as in section VIC for the fluid phase. We note that
for the stability matrix sector
of moments unequal to zero there are only eigenvalues
larger than zero
\cite{Carlucci1}.

Finally, we sketch the proof given in Ref.\,\onlinecite{Carlucci1}
that eigenvalues of the first two families, which is the longitudinal
 sector and the anomalous one, has only eigenvalues
which are larger or equal to the
replicon eigenvalues given above.
With the definitions
\begin{equation}
\Delta_{r}^{k}\equiv  \left\{
\begin{array}{ccc}
\frac{1}{2} (p_r-p_{r+1}) & \mbox{for} & r < k-1   \\
\frac{1}{2}(p_{k-1}-2 p_k) & \mbox{for} & r=k-1   \\
p_r- p_{r+1} & \mbox{for} & r > k-1  \,.
\end{array} \right. \label{a50}
\end{equation}
and
\begin{equation}
\Lambda_{k}(r)\equiv  \left\{
\begin{array}{ccc}
 \lambda(r;k, r+1)   & \mbox{for} & k \ge  r+1    \\
  \lambda(r;r+1,r+1)       & \mbox{for} & r > k-1  \,.
\end{array} \right.       \label{a60}
\end{equation}
we obtain
\begin{equation}
\rm{det}\left( \underline{M}^{(k)} - \lambda I \right)  =
\prod_{r=0}^{R}
\left[
 \Lambda_k(r)
 -
 \lambda
 \right]
 \rm{det}
\left[I+ {\underline{M}'}^{(k)} \right]   \label{a70}
\end{equation}
with
\begin{equation}
{M'}^{(k)}_{rs} =
K_k^{rs} \frac{\Delta_{s}^{k}}{2 (\Lambda_k(s)-\lambda) }   \label{a80}
\end{equation}
where we denote $ {\underline{\rm M}'}^{(k)} $ by the matrix
$ {M'}^{(k)}_{rs} $
and  $ K_k^{rs} $ is a generalized discrete Fourier transform of the
ultrametric matrix $ \underline{\rm M}^{(k)} $ \cite{Carlucci1}.
The matrix $ K_k^{rs} $ is denoted as the kernel for the ultrametric matrix
$ M^{\alpha \beta, \gamma \delta} $ which means  that $
M^{\alpha \beta, \gamma \delta}$ is given by
$ M^{(k)}_{rs}= \Lambda_k(r)+K_k^{rs} \Delta_{s}^{k}/2 $
in the longitudinal or anomalous sector $k $ and $r $, $s $ runs over the
basis vectors in the $ k$ sector.

We point out that $ K_k^{rs} $ can be written as
\begin{equation}
 K_k^{rs}= 4 B_k({\rm max}(r,s))    \label{a85}
\end{equation}
 where
the function $ B_k $ can be expressed explicitly by the Green functions
$ G_s $ \cite{Carlucci1} so does not depend on the disorder 
function $ f $. One finds
\begin{equation}
 B_k(r) < 0  \qquad   \mbox{and}  \qquad  B_k(r+1) -B_k(r)>0  \,.
\label{a87}
\end{equation}
Denoting $ \rm{det}_{{\cal S},{\cal S}'}[{\underline{\rm M}'}^{(k)}] $
by the determinant of the sub-matrix of $ {\underline{\rm M}'}^{(k)} $
with lines in $ \quad {\cal S} \in \{0, \ldots, R\} $ and columns in
 $ \quad {\cal S}' \in \{0, \ldots, R\} $ where we suppose that $ {\cal S} $
 and $ {\cal S}' $ has the same number of elements denoted by
$ \#{\cal S}'=\#{\cal S}' $.
We obtain
\begin{equation}
\rm{det}\left( \underline{M}^{(k)} - \lambda I \right) =
\prod_{r=0}^{R}
\left[
 \Lambda_k(r)
 -
 \lambda
 \right]
 \sum_{\cal S} {\rm det}_{{\cal S},{\cal S} }[{\underline{M}'}^{(k)}] \,.
\label{a90}
\end{equation}
In \cite{Carlucci1} it is shown that the right hand side is larger than zero
for $ \lambda \le {\rm Min}_r[\Lambda_k(r)] $. This means
that the eigenvalues in the
non-replicon sectors are always larger than the smallest eigenvalue in the
replicon sector. We now give a more general proof of this fact useful in the
next subsection:

This is true if $ {\rm det}_{{\cal S},{\cal S}'}
[{\underline{\rm M}'}^{(k)}] \ge 0 $
for  $ \quad {\cal S}, {\cal S}'\in \{0, \ldots, R\} $.
We suppose the ordering
$ s_{i} < s_{i+1} $ for $ s \in {\cal S} $ and similar for $ {\cal S}' $.
By subtracting appropriate line and columns of the matrix
$ \underline{\rm K}_k $ where
$ \underline{\rm K}_k $ denotes  the matrix $ K^{rs}_k $
 we obtain a matrix where its determinant is
built purely from its diagonal elements given by
\begin{align}
&  {\rm det}_{{\cal S},{\cal S}'}[\underline{\rm K}_k ] =
4^{\#{\cal S}} B_k({ \rm max}( s_{\#{\cal S}} ,
s'_{\#{\cal S}'}) )  \nonumber \\
&
\times \prod_{i=1}^{\#{\cal S}-1}
\left[B_k({ \rm max} (s_i,s'_i) )-B_k({ \rm max}( s_{i+1},
  s'_{i+1}))\right]   \,.    \label{a95}
\end{align}
By using (\ref{a80}) with $ \Delta^k_r \le 0 $ (\ref{a50}) for $ n \to 0 $ 
and (\ref{a87})
we obtain $ {\rm det}_{{\cal S},{\cal S}'}
[{\underline{\rm M}'}^{(k)}] \ge 0 $ for $ \lambda \le
{\rm Min}_r[\Lambda_k(r)] $.

It is clear from the considerations above that the various sectors
especially the longitudinal sector depends on the concrete
hierarchical structure we choose. This means, that we can also
get other eigenvalues
for the various sectors by starting from a given minimal level
of hierarchy  by an appropriate
artificial division of the various sectors leading to a larger level of
hierarchy. It is clear that nevertheless the lowest eigenvalue
of the stability matrix being in the replicon sector didn't change.
Now suppose, we try to restrict the stability matrix $ M^{\alpha \beta,
\gamma \delta} $ to the $ k=0 $ longitudinal sector of
a suitable subdivided hierarchy, corresponding to a search of
the minimum of $ F_{\rm var} $ (\ref{115}) in the self-energy matrices
$ \sigma_{\alpha \beta} $ which are contained in the Parisi-algebra.

For a subdivision of blocks we obtain that $ B_k(r) $ given explicitly in
\cite{Carlucci1} is constant on two blocks in the subdivided hierarchy
originating  from the same blocks $ k $ and $ r $
in the precursor hierarchy. Furthermore, we get doublings in the eigenvalues
$ \Lambda^{k}(r) $
corresponding to the subdivision.
But this results to $ {\rm det}_{\cal S, \cal S
 }[\underline{\rm M}^{(k)}] \not=0  $
only if  $ {\cal S } $ does not contain two blocks in the subdivided hierarchy
originating from the same block.
Then we immediately obtain
from (\ref{a80}), (\ref{a85}), (\ref{a90}) that we can always
subdivide the hierarchy in such a way that the lowest eigenvalue
in the $ k=0 $ longitudinal sector is given by the minimum of the eigenvalues
in the replicon sector $ {\rm Min}_r[\Lambda_0(r)] $. This means that by
restricting the stability matrix $ M^{\alpha \beta, \gamma \delta} $
to the subspace of symmetric
self-energy matrices in the Parisi-algebra with the constraint (\ref{165})
we obtain
that the lowest eigenvalue of the restricted matrix
$ M^{\alpha \beta,\gamma \delta} $ is equal to the lowest
eigenvalue in the replicon sector.

\subsection{Stability in the
variational approach of M\'ezard  and Parisi}
In this subsection, we carry out a similar analysis for the Bogoliubov
variational approach of the M\'ezard-Parisi theory,
outlined in Section IV, as was done for the
large $N'$-limit theory in the last subsection. The self-energy within
this approach is calculated by searching for the stationary points of
the variational free energy (\ref{110}). We get a stability matrix
of this stationary point by taking the second derivative of
$ F_{\rm var} $ with respect to the self-energy matrix
under the constraint (\ref{165}).
This was calculated in Ref.\,\onlinecite{Sasik1}. We obtain
\begin{eqnarray}
 \tilde{M}^{\alpha \beta, \gamma \delta} & = &
\frac{K_B T}{N v^2}
\frac{\partial^2 F_{\rm var} }{\partial \sigma_{\alpha \beta}
\partial \sigma_{\gamma \delta} }        \label{a200}  \\
& = & \frac{1}{2} L^{(2)}_{\alpha \beta,\gamma \delta}
 - L^{(2)}_{\alpha \beta, \alpha' \beta'}
\tilde{f}''(L^{(1)}_{\alpha' \beta', \gamma' \delta'})
L^{(2)}_{\gamma' \delta', \gamma \delta}            \nonumber
\end{eqnarray}
This matrix corresponds to the matrix $
 M^{\alpha \beta, \gamma \delta}  $ (\ref{a20})  in the
large $ N' $-limit approach.
Because $ \tilde{M}^{\alpha \beta, \gamma \delta}$
 is a
ultrametric matrix we obtain by using the rather general consideration
for eigenvalues in  the {\it replicon sector}  of
these types of matrices \cite{Temesvari1}
\begin{equation}
\tilde{\lambda}(r; k,l)=
 L'_{kl}\left\{ \frac{1}{2}
 - \tilde{f}''(2 (K_B T/v)[g_{R+1}-g_r)] L'_{kl} \right\}\,.  \label{a210}
 \end{equation}
By comparing (\ref{a210}) with (\ref{a40})
we obtain also in the variational approach that the most
divergent eigenvalues $ \tilde{\lambda}(r; r+1,r+1) $ are zero in the
continuous replica symmetry-breaking solutions
as was also the case in the large
$ N' $-limit approach.

In the following we show  that in
the variational approach
the eigenvalues of the longitudinal and anomalous sectors are larger than zero.
We first define the reduced stability matrix
\begin{equation}
 \underline{\tilde{\rm M}}_{\rm red} \equiv
(\underline{\rm L}^{(2)})^{-1} \underline{\tilde{\rm M}}
(\underline{\rm L}^{(2)})^{-1}      \label{a215}
\end{equation}
Here $ \underline{\rm L}^{(2)} $ is the matrix $ L^{(2)}_{\alpha \beta,
\gamma \delta} $.
By using that the kernel of $  \underline{\rm L}^{(2)} $ is given by
 $ -K_k^{rs} $ we obtain for the kernel of $ (\underline{\rm L}^{(2)})^{-1} $
\cite{Carlucci1}
\begin{equation}
 \underline{\rm F}^0_k  =  (\underline{\Lambda}^0_k)^{-1/2}
(\underline{\rm N}^0)_k^{-1} (\underline{\Lambda}^0_k)^{-1/2} \;
\underline{\rm K}_k  (\underline{\Lambda}^0_k)^{-1}
\label{a216}
\end{equation}
with
\begin{equation}
\underline{\rm N}^0_k =
{\rm I}- \frac{1}{2}
(\underline{\Lambda}^0_k)^{-1/2}
\underline{\rm K}_k
\underline{\Delta}^k
(\underline{\Lambda}^0_k)^{-1/2}
  \label{a217}
\end{equation}
where
$ (\underline{\rm \Lambda}_k^0)_{rs}=\Lambda^0_{k}(r) \, \delta_{rs} $
and $ \Lambda^0_k(r) $ is given by
\begin{equation}
\Lambda^0_{k}(r) \equiv  \left\{
\begin{array}{ccc}
 L'_{k, r+1}   & \mbox{for} & k \ge  r+1    \\
 L'_{r+1,r+1}       & \mbox{for} & r > k-1  \,.
\end{array} \right.       \label{a220}
\end{equation}
We mention that $ \underline{\rm L}^{(2)} $ has
positive eigenvalues which can be seen from the
positivity of $ G^{-1}_{\alpha \beta} $. This has to be assumed
for the stability of    $ F_{\rm trial} $,   leading  to the positivity of
$ \underline{\rm L}^{(2)}  $ for $ n \to 0 $ because only
the first term in (\ref{a200}) is unequal
to zero for $ \tilde{f}=0 $ where the second term in
(\ref{125}) does not contribute to the stability matrix \cite{Sasik1}.
Now we use that $ \underline{\rm L}^{(2)} $ is given by
$ \underline{\Lambda}^0_k-\underline{\rm K}_k \underline{\Delta}^k/2 $
which means that
$ \underline{\rm N}^0_k$ has only positive eigenvalues.
By using the same considerations for the eigenvalue equation
$ {\rm det}[\underline{\rm N}^0_k -\lambda] =0 $ as was done at
the end of the last subsection we obtain further
that all eigenvalues of
$ \underline{\rm N}^0_k$
are lower than one.
This leads to the fact that the denominator
 $ (\underline{\rm N}^0_k)^{-1} $ in
$ \underline{ \rm F}^0_k $ (\ref{a216})
can be expanded in a geometric series.

The eigenvalue equation
for $ \underline{\tilde{\rm M}}_{\rm red } $ is given by the right hand side
of Eq.~(\ref{a70})
with $ K_k^{rs} $ in (\ref{a80}) is substituted by
the expanded form of $ (F_k^0)^{rs} $ (\ref{a216}).
$ \Lambda_k(r) $ is built of the replicon eigenvalues
of $\underline{\tilde{\rm M}}_{\rm red } $  corresponding to (\ref{a60}).
By carrying out the calculation of the resulting
sub-determinants of sums and products of matrices
by standard rules (Cauchy-Binet formula) we obtain as in the last subsection
that the non-replicon eigenvalues of $ \underline{\tilde{\rm M}}_{\rm red} $
are larger than the
smallest replicon eigenvalue given in (\ref{a210}).

Furthermore, we  obtain also with a similar proof as in
subsection 1 that the projected matrix
$ \tilde{M}^{\alpha \beta, \gamma \delta}_{\rm red} $ to
the space of the symmetric self-energy
matrices $ \sigma_{\alpha \beta} $ of the Parisi form
with the constraint (\ref{165}) contains the
smallest replicon eigenvalue.

Up to now, we have only shown that the results of the large $ N' $ approach
considered in the last subsection are also valid
for the reduced stability matrix
$ \tilde{M}^{\alpha \beta, \gamma \delta}_{\rm red} $.
It is not clear whether this is also valid for the full stability
matrix $ \tilde{M}^{\alpha \beta, \gamma \delta} $ (\ref{a200}) of the
M\'ezard-Parisi variational approach.
Nevertheless, one normally does not need the results above
in their general form for a stability analysis of saddle point solutions
of (\ref{115}). It is enough for this analysis to know
the results concerning the positivity of the eigenvalues.
This can be immediately reached by using the defining equation  (\ref{a215})
of the reduced stability matrix and the general conclusions above.

This leads to the following results for the stabilities in
the large $ N' $ and the variational approach of the
M\'ezard-Parisi theory:
\begin{enumerate}
\item
The eigenvalues in the replicon sector
are given by $ \lambda(r;k,l) $ in (\ref{a40}) for the
large $ N' $-approach and by $ \tilde{\lambda}(r;k,l) $
in (\ref{a210}) for the variational approach.
In the case that the eigenvalues in the replicon sector are all positive
in the large $ N' $-approach or the variational approach
we obtain also that all eigenvalues of the full stability matrices
$ M^{\alpha \beta, \gamma \delta} $ or
$ \tilde{M}^{\alpha \beta, \gamma \delta} $, respectively, are
larger than zero. This leads to the stability of the
corresponding saddle point solution.
\item
The eigenvalues of the continuous symmetry-breaking solution are larger
than or equal to zero.
\item
The eigenvalues of the
stability matrix projected on the subspace of variations in the
symmetric self-energy
matrices $ \sigma_{\alpha \beta } $ in the Parisi algebra
with the constraint (\ref{165}) are larger than
or equal to zero
if and only if the eigenvalues of the full stability  matrix
not restricted to variations in the Parisi algebra in both approaches
are larger than or equal to zero. \\[0.2cm]
\end{enumerate}
We further note that the eigenvalues $ \lambda(r;k,l) $ (\ref{a40}) and
$ \tilde{\lambda}(r;k,l) $ (\ref{a210}) of both approaches are proportional
to each other with a positive proportional constant
when neglecting the distinction  in the effective disorder functions
$ f $ and $ \tilde{f} $ related by (\ref{a10}).

\section{Instability of finite-step
replica symmetry-breaking solutions
in the fluid phase}
In this section we  show in general that finite-step
replica symmetry-breaking solutions
for the fluid phase of the vortex lattice
with a Gaussian disorder correlation function (\ref{15}) are not stable.
This will be shown
irrespective of the number of steps. We have shown this in
the case of one-step replica symmetry breaking in Section~VI.
From (\ref{170}) and (\ref{250}) we obtain in the case of
a $ R $-step replica symmetry-breaking solution in the fluid phase
\begin{eqnarray}
& \Delta f_{\rm var}=\displaystyle\frac{k_B T }{2} \sum_{i=1}^{R} \bigg\{
 \left[\frac{1}{m_{i+1}}-\frac{1}{m_i}\right] S(\tilde{\Delta}_{m_i})
 \nonumber \\
& + [m_{i+1}-m_i] \; D(\,2B[\Delta_{m_i}])\bigg\}           \label{a500}
\end{eqnarray}
with
\begin{eqnarray}
S(x) & = & \frac{1}{2} \left[4 \rm{arcsinh}\left(\frac{x^{1/2}}{2}\right)-
\frac{x^{1/2}}{(1+x/4)^{1/2}}\right] \,,  \label{a510} \\
B[\Delta_{m_i}] & = & \frac{k_B T}{v}  \bigg\{\sum_{j=i}^{R-1}
\frac{1}{m_{j+1}}\left[g(\Delta_{m_j})-g(\Delta_{m_{j+1}})\right]
 \nonumber \\
 & & + g(\Delta_{R})\bigg\} \,.
\label{a520}
\end{eqnarray}
In (\ref{a500}) we used that $ \Delta_0=0 $ and $ m_{R+1} \equiv 1 $.
The $ R $ stationarity conditions
$ \partial \Delta f_{\rm var}/\partial \Delta_i =0  $ for
$ i=1 \ldots R $ lead to
\begin{equation}
\sum_{i=1}^l
\frac{\Delta_{m_i} \! - \! \Delta_{m_{i-1}}}{m_i} \! =
\!- 2 \frac{k_B T}{v} D'(2B[\Delta_{m_l}]).
 \label{a530}
\end{equation}
for $ l=1 \ldots R$ corresponding to (\ref{253}) in the case of the one-step
replica symmetry-breaking solution. The saddle point conditions
$ \partial \Delta f_{\rm var}/\partial m_i =0  $ for
$ i=1 \ldots R $ lead to
\begin{equation}
\sum_{i=1}^l \frac{1}{m^2_i} \left[S(\tilde{\Delta}_{m_i})-
S(\tilde{\Delta}_{m_{i-1}})\right]
= {\cal D}(\,2B[\Delta_{m_l}])- Z_l  \label{a540}
\end{equation}
for $ l=1,\ldots,R $ with
\begin{eqnarray}
 Z_l  & = &-
 \sum_{j=1}^l \sum_{i=1}^{j-1} \frac{\Delta_{m_i}}{m_j}
\left(\frac{1}{m_i}-\frac{1}{m_{i+1}} \right)
\nonumber \\
 & & \times \left[g(\Delta_{m_j})-g(\Delta_{m_{j-1}})\right]\ge 0  \,.
 \label{a550}
\end{eqnarray}
One should compare this equation with (\ref{255}) in the case of
a one-step replica symmetry-breaking solution.
Since $ Z_1=0 $ we can use (\ref{a530}) and (\ref{a540}) similarly as in
the derivation of (\ref{840}) to obtain $ \tilde{\lambda}(1;2,2) < 0 $
irrespective of the number $ R $
of hierarchical steps for $ \Delta_1 > 0 $. Furthermore,
we obtain $ \tilde{\lambda}(1;2,2) = 0 $
at $ \Delta_1=0 $ as was also the case in the one-step hierarchical
symmetry-breaking case.
\end{appendix}

\end{document}